\documentclass[final,1 p,times]{elsarticle}

\usepackage{amssymb}
\usepackage{amsmath}

\usepackage{pifont}
\usepackage{multicol}
\usepackage{tcolorbox}
\usepackage{listings}
\usepackage{xcolor}
\usepackage{subcaption}
\usepackage{listings}
\usepackage{tabularx}

\definecolor{shadecolor}{gray}{0.9}
\usepackage{framed}
 {\endMakeFramed}

\lstdefinestyle{python}{
    language=Python,
    basicstyle=\ttfamily\scriptsize,
    keywordstyle=\color{blue}\bfseries,
    stringstyle=\color{green!50!black},
    commentstyle=\color{gray},
    numbers=left,
    numberstyle=\scriptsize\color{gray},
    showstringspaces=false,
    frame=single,
}

\lstdefinelanguage{Alloy}{
    keywords={sig, fact, pred, all, some, no, one, disj, set, enum},
    sensitive=true,
    morecomment=[l]{--},
    morestring=[b]{"},
}
\lstdefinestyle{alloy}{
    language=Alloy,
    basicstyle=\ttfamily\scriptsize,
    keywordstyle=\color{purple}\bfseries,
    stringstyle=\color{red!50!black},
    commentstyle=\color{gray},
    numbers=left,
    numberstyle=\scriptsize\color{gray},
    showstringspaces=false,
    frame=single,
}

\lstdefinelanguage{OCL}{
    keywords={context, inv, pre, post, let, if, then, else, endif, and, or, not},
    sensitive=true,
    morecomment=[l]{--},
    morestring=[b]{"},
}
\lstdefinestyle{ocl}{
    language=OCL,
    basicstyle=\ttfamily\scriptsize,
    keywordstyle=\color{teal}\bfseries,
    stringstyle=\color{orange},
    commentstyle=\color{gray},
    numbers=left,
    numberstyle=\scriptsize\color{gray},
    showstringspaces=false,
    frame=single,
}

\usepackage{booktabs}
\usepackage{multirow}
\usepackage{color}
\usepackage{framed}

\usepackage[normalem]{ulem} 
\usepackage{xcolor}

\newcommand{\ins}[1]{#1} 
\newcommand{\del}[1]{} 
\newcommand{\chg}[2]{#2} 

\journal{Journal of Systems and Software}

\begin{document}

\begin{frontmatter}



\title{A framework for assessing the capabilities of code generation of constraint domain-specific languages with large language models}


\author{David Delgado} 
\affiliation{organization={Universitat Oberta de Catalunya},
            addressline={Rambla del Poblenou, 154-156}, 
            city={Barcelona},
            postcode={08018}, 
            state={Barcelona},
            country={Spain}}

\author{Lola Burgueño} 
\affiliation{organization={ITIS Software, Universidad de Málaga},
            addressline={Avda. Cervantes 2}, 
            city={Málaga},
            postcode={29071}, 
            state={Málaga},
            country={Spain}}
\author{Robert Clarisó} 

\affiliation{organization={Universitat Oberta de Catalunya},
            addressline={Rambla del Poblenou, 154-156}, 
            city={Barcelona},
            postcode={08018}, 
            state={Barcelona},
            country={Spain}}

\begin{abstract}
%
%
%
%
%
%
%
%
%
Large language models (LLMs) \chg{are}{can be} used to support software development tasks, \emph{e.g.}, through code completion or code generation. However, their effectiveness drops significantly when considering less popular programming languages such as domain-specific languages (DSLs). 

\ins{In this paper}, \chg{W}{w}e propose a generic framework for \chg{the evaluation of}{evaluating} the capabilities of LLMs generating DSL code from textual specifications\chg{, in particular,}{. The generated code is assessed} from the perspectives of \del{the} well-formedness and correctness\del{of the code}. 
\chg{We assess the effectiveness of our framework by studying the capabilities 
of different LLMs for code generation tasks for}{This framework is applied to} a particular type of DSL\chg{:}, constraint languages\chg{. In particular, we have evaluated}{, focusing our experiments on} OCL and Alloy \ins{and comparing their results to those achieved for}\del{.
We study the performance of LLMs on 28 domains and compare the outcomes for OCL and Alloy, contrasting these results to those for} Python, a popular general-purpose programming language. 

\chg{Our experiments}{Experimental results} show that\chg{ (1)}{,} in general, LLMs have better performance for Python than for \del{the constraint DSLs} OCL and Alloy.
\ins{LLMs with smaller context windows such as open-source LLMs may be unable to generate constraint-related code, as this requires managing both the constraint and the domain model where it is defined.
Moreover, some improvements to the code generation process such as code repair (asking an LLM to fix incorrect code) or multiple attempts (generating several candidates for each coding task) can improve the quality of the generated code. Meanwhile, other decisions like the choice of a prompt template have less impact. All these dimensions can be systematically analyzed using our evaluation framework, making it possible to decide the most effective way to set up code generation for a particular type of task.}

\del{;(2) our framework is effective to study the capabilities of different LLMs when generating code for constraint DSLs; (3) the LLM of choice affects the quality of the results, with code repair and multiple attempts significantly improving code quality while the prompting template having a lesser impact.
The framework proposed in this paper enables an effective evaluation of LLMs for generating DSL code. In particular, after the application of the framework we can conclude that constraint languages offer a challenge from a code generation perspective, as it is necessary to encode the schema where the constraints are defined.}


\end{abstract}





\end{frontmatter}




\section{Introduction}
\label{sec:introduction}

Recent advances in generative artificial intelligence (AI), and in particular \emph{large language models (LLMs)}, have made significant progress towards generating source code from textual requirements. Such tools can be used to improve developer productivity~\cite{murali_ai-assisted_2024}, \emph{e.g.}, integrating code autocomplete tools in integrated development environments (IDEs). Moreover, they facilitate the rapid exploration of ideas or the quick creation of prototypes supported by AI tools, \emph{e.g.}, AI software engineering agents~\cite{rasheed_autonomous_2024} and the notion of \emph{vibe coding\footnote{Vibe coding is an approach to software development where the developer collaborates with AI tools to brainstorm, code, test and debug.}}. 

%

The power of LLMs comes from the vast dataset used in its training phase. As a result, when it comes to code generation, LLMs are able to extract patterns from a large collection of code samples, effectively learning the syntax and semantics of programming languages and the relationships between textual requirements (\emph{e.g.}, comments in source code) and their source code implementation. Nevertheless, this is only the case for popular \emph{general-purpose} programming languages (GPL), such as Python, Java, C or C\#, where there is a large collection of samples publicly available. The performance of LLMs worsens when dealing with \emph{low-resource programming languages} such as \emph{domain-specific languages} (DSLs)~\cite{cassano_knowledge_2024}. In these languages, the number of available code samples is small compared to GPLs. The lack of examples makes the LLM's knowledge of the language less precise.
Therefore, evaluating the quality of the code generated by LLMs is critical, especially for DSLs.

A notable subset of domain-specific languages (DSLs) includes formal constraint and query languages, such as OCL~\cite{cabot_object_2012} and Alloy~\cite{jackson_software_2006}. Unlike natural language, which is inherently ambiguous and imprecise for specifying system behavior, these languages provide a precise and formal notation~\cite{bajwa_ocl_2010}. Thus, they are particularly well-suited for expressing integrity constraints in software systems, such as invariants, pre-conditions and post-conditions. These constraints support rigorous verification, validation, and testing. However, these languages create several unique challenges from the point of view of code generation: 
\begin{itemize}
    \item They are low-resource programming languages, which makes LLMs perform worse than for general-purpose languages.
    \item The constraints or queries they define refer to a \emph{schema} or specification in which the constraint should be checked or enforced. That is, code generation needs to manage two different tasks at once: how the schema is provided (or generated) and how the constraints use the information therein. \item They are declarative rather than procedural. Hence\ins{,} it is not always possible to \emph{execute} them, but they need to be verified or tested against positive and negative samples. 
    \item Constraints are usually \emph{global} rather than \emph{local}. As result, it is harder to consider them in isolation, as they may interact among them and cause unintended side-effects. This makes it harder to consider requirements individually as sometimes several constraints need to be dealt simultaneously to reach a satisfactory solution.
\end{itemize}


In this paper, we present a modular framework for assessing LLM-based code generation, which supports both GPLs and constraint and query DSLs. The framework supports different prompt styles, prompt augmentation techniques and quality evaluation strategies. 
This framework enables the exploration of different alternatives for code generation, which allows developers to identify the best approach to generate code for their particular language.
To validate the effectiveness of our framework, we have instantiated it for two DSLs (Alloy and OCL) and one GPL (Python).
For our experiments, we have collected and curated a set of examples in OCL, Alloy and Python composed of domain specification plus integrity constraints. 
We have instantiated our framework and used it to evaluate the code generation capabilities of four different LLMs, namely DeepSeek-coder, GPT-4o, GPT-4o-mini, and Llama 3.1.

\ins{Specifically, this paper presents the following novel contributions to the field of LLM-based code generation:}
\begin{enumerate}
\item \ins{A flexible and highly configurable modular assessment framework for code generation in which all relevant decisions about the code generation process (prompt template, LLM, use of code generation, use of multiple attempts, \ldots) can be parameterized. In this way, it is possible to study the impact of decisions on the code generation process. While other works on code generation focus on one or a few dimensions, this framework can adapt to diverse code generation scenarios and supports the systematic comparison of multiple code generation settings.}
\item \ins{A study of the specific challenges surrounding code generation for the constraint languages OCL and Alloy, comparing the results to those achieved for general-purpose languages like Python, and enriching existing datasets with additional synthetic data.}
\item \ins{A set of comprehensive code generation experiments that explores and evaluates all possible configurations (more than 90k) for a set of coding tasks in order to select the best possible alternative.}
\end{enumerate}


The remainder of this paper is organized as follows. Section \ref{sec:background} presents the necessary background to contextualize our approach. Section \ref{sec:framework} introduces our modular LLM-based  evaluation framework, while Section~\ref{sec:instantiation} describes the particular instantiation of the framework that we use in the experiments in this paper. 
In Section \ref{sec:case-study}, we illustrate the application of this framework and assess the quality of the generated code through several experiments. Section \ref{sec:related-work} discusses related work in the field, positioning our contribution within the broader research landscape. Finally, Section \ref{sec:conclusions} concludes the paper and outlines directions for future research. 

\section{Background}
\label{sec:background}

Since their inception, LLMs have been used for a variety of code-related tasks, such as code completion (\emph{i.e.}, proposing code snippets that extend a code fragment), automatic test generation, and generation of code from textual comments. In this paper, we focus on the latter: the generation of code from textual specifications. In this section, we present the notion of code generation, how it relates to DSLs and techniques used to improve the performance of LLMs in this task.


\subsection{Code generation}

In the context of software engineering, the term \emph{code generation} has been used with a variety of meanings. For instance, in the context of model-based software engineering (MBSE), code generation usually refers to the translation of a high-level formal specification (a  model\footnote{Note that the term ``model'' is used with different meanings in AI and MBSE. In the AI field, a model is a mathematical representation of the patterns learned from a dataset, \emph{e.g.}, a trained neural network or a statistical model. Meanwhile, in MBSE a model is an abstraction of a system or software artifact. In this paper, we use a particular type of software model, which we call interchangeably \emph{domain} or \emph{conceptual model}.
To avoid confusion, each time we talk about an AI model, we will refer to it with a precise name, \emph{e.g.}, AI model, machine learning model, language model\chg{ ,}{or} LLM} of a software system) into low-level executable code. In this scenario, the formal nature of the input notation enables the use of dedicated model transformation and templates, which use a combination of patterns and rules to match elements in the original specification and declare their intended translation.


On the other hand, code generation consists in the creation of source code from a flexible high-level description in natural language or partial code snippets~\cite{chen_evaluating_2021}. This is the definition of code generation that we adopt in this paper. This type of code generation helps translate human intent into machine-executable code, allowing developers to quickly prototype, automate repetitive tasks, and even generate complex software systems.

Code generation has undergone a significant transformation since the appearance of LLMs, achieving remarkable capabilities in understanding and generating code in various programming languages~\cite{chen_evaluating_2021, murali_ai-assisted_2024}. LLMs are already pre-trained with a vast amount of data, including code written in several programming languages and even DSLs. Consequently, there is no need to train them to achieve positive results in code generation tasks. However, not every language is equally represented in the pre-training dataset of an LLM, making the effectiveness and quality of code generation with LLMs language-dependent.

\subsection{Low-resource programming languages and DSLs}


The effectiveness of AI methods based on machine learning depends on the quality and quantity of data available for training. Having less data \del{(}or worse data\del{)} reduces the quality of the patterns discovered in the learning process: they may be incorrect, \emph{e.g.}, assume a particular case can be generalized; or incomplete, \emph{e.g.}, they may fail to include unusual corner cases.

In the context of natural language processing, \chg{\emph{e.g.},}{such as} in machine translation, languages are typically classified as \emph{high resource} if there are many publicly available \ins{high-}quality samples that can be used for training purposes\ins{,} and \emph{low resource} if there are only \ins{a} few.
In code generation, this problem is also discussed as the high, low, and very low resource programming languages problem \cite{roziere_code_2024, guo_deepseek-coder_2024,li_starcoder_2023, cassano_multipl-e_2023, cassano_knowledge_2024, joel_survey_2024}. Popular general-purpose languages tend be high resource, while  DSLs tend to be at the lower end of these classifications, with few available examples. The scarcer the data about a language, the more difficult it is for an LLM to generate code that follows the syntax and semantics of the language and implements the input specification faithfully.


DSLs present additional challenges in code generation with LLMs, making it even more necessary to adjust the code generation process. First, there is a lack of benchmarks \cite{joel_survey_2024} and training datasets \cite{pan_generative_2024, abukhalaf_pathocl_2024}, which makes it challenging even to generate code that can be successfully parsed. Moreover, there are fewer tools available for code evaluation, making it necessary to create language-specific tools to check if the generated code achieves the intended goals. 

\subsubsection{Prompt engineering}
Prompting is a primary method for interacting with LLMs. With this technique, the user inputs a set of instructions in natural language (the prompt), asking the LLM to generate a response that aligns with the given information. 

The construction of effective prompts, called \emph{prompt engineering}, is a rapidly evolving field of study. There are many different prompt engineering techniques, some of which can be applied to any kind of task and others which are specific to particular tasks. Nevertheless, not every technique works for every use case and not every technique works with any LLM size~\cite{wei_emergent_2022}. There are several techniques that are worth mentioning in the context of this paper:
\begin{itemize}
\item Chain of Thought (CoT) is used to deal with complex tasks by decomposing them into simpler steps to guide the LLM \cite{wei_chain--thought_2022}. The entire reasoning process occurs within a single prompt, and the LLM generates the entire chain of reasoning in one go. 
\item Iterative prompting (or dialogue-based prompting, or conversational prompting) consist of prompting the LLM, receiving the response and incorporating it into the context of a new prompt. In this case, the reasoning spreads across multiple interactions and may involve external interactions (such as a tool invokation) and decision-making.
\end{itemize}

\emph{Prompt augmentation}, also called \emph{context engineering}, is used to provide more information or context to a base prompt. This additional information is always relevant to the task at hand and helps the LLM improve the quality of its response~\cite{wei_emergent_2022}. An example of relevant context in the context of this paper would be the description of the domain where the constraints are defined.

Although prompt augmentation is limited by the maximum context window size of LLMs, this shortcoming can be avoided by carefully selecting which information is provided as context. That is, considering the task being addressed and the prompt, we select the subset of fragments from the context that is relevant from our particular task. This technique, called \emph{retrieval-augmented generation} (RAG) has been used to help the LLM adhere to the grammar of a specific target language, to improve the its ability to generate code that is syntactically valid~\cite{bassamzadeh_plan_2024}.

At the same time, LLMs are in-context learners, this is, they demonstrate the ability to acquire new skills directly from prompt interactions, particularly through a limited number of examples~\cite{brown_language_2020}. This capability allows us to define a new classification and categorize prompting strategies depending on the number of examples provided within the prompt: \emph{zero-shot prompting}, where the LLM performs a code task without any explicit examples, and \emph{few-shot prompting}, where the LLM learns from a small set of examples included in the prompt.




\del{Note that all the prompt engineering techniques presented so far can be used in isolation or in combination with others.}

\subsection{Evaluation of generated code}
\label{sec:rel-eval}

LLMs can produce code that is not syntactically valid or that does not accurately implement the provided requirements. Thus, it is important to assess the quality and suitability of the generated code. There are different ways to evaluate LLM-produced code. In this sub-section, we discuss the most relevant approaches and, particularly, those used in the remainder of this paper.

A straight-forward way to evaluate generated code is its \emph{manual} inspection carried out by human experts. Unfortunately, manual inspection can be very time-consuming. LLMs can generate code much faster than its manual review takes. As a result,  manual inspection is not feasible in scenarios where code generation is used intensively or in large software systems. To address this challenge, automated evaluation metrics have been proposed.


Automatic code evaluation can be split into two sub-groups: (1) assessing whether the code is syntactically and semantically valid in the given programming language (\emph{i.e.}, the code is \emph{well-formed}), and (2) assessing whether the generated code fulfills the intended functionality as specified by the user (\emph{i.e.}, the code is  \emph{correct}).

A piece of code is considered syntactically valid or \emph{well-formed} when it follows the basic structural and syntactic rules of the language, \emph{i.e.}, it can be parsed without errors. Well-formedness can be evaluated using a parser for the target language.

Moreover, a piece of code is considered semantically valid when it is well-formed and conforms to specific semantic rules, \emph{e.g.}, not allowing the use of undefined variables. Semantic validation can be evaluated by using a semantic analyzer, \chg{that it}{which} is usually part of the language compiler. Nevertheless, it is often assessed implicitly by code execution \cite{chen_evaluating_2021, olausson_is_2024}. As a consequence, it is usually a part of correctness evaluation, which also involves code execution.

\emph{Correctness} is the term used to evaluate whether a piece of code meets the intended specification~\cite{olausson_is_2024, hou_large_2024}.
One strategy to assess correctness is considering it a matching problem: look for an exact match of the generated code with a provided solution that it is known as correct. However, this approach has limited applicability due to the expressiveness of programming languages: there are many different ways to write code that has exactly the same behavior.
The number of alternative solutions to a coding task is so large that in practice it is impractical to assess correctness by looking for matches against a reference solution~\cite{dong_codescore_2025}. A similar problem appears if, instead of using exact matches, we rely on similarity metrics from natural language processing such as BLEU or ROUGE~\cite{hou_large_2024}. Again, these approaches are adequate for machine translation and text summarization, but are inadequate for code evaluation~\cite{liu_is_2023,ren_codebleu_2020}, due the expressiveness of programming languages.


Another strategy is assessing functional correctness by means of unit tests~\cite{chen_evaluating_2021, mathews_test-driven_2024}. In this approach, the generated code is considered correct only if it passes the unit tests. Moreover, functional correctness is assessed predominantly using two metrics: \emph{accuracy} and \emph{pass@k}~\cite{olausson_is_2024, rasheed_large_2024}. Accuracy measures the number of times that  generated code is correct in the first attempt. On the other hand, pass@k  measures the likelihood that at least one of the \textit{k} generated code samples achieves functional correctness \cite{kulal_spoc_2019}. This metric not only indicates how often the model succeeds in the first attempt, considering that pass@1 is equal to accuracy, but also highlights the potential benefit of multiple generation attempts \cite{olausson_is_2024}. 

Last but not least, large language models have been recently integrated into evaluation processes, leading to the emergence of the ``LLM-as-a-judge'' paradigm \cite{sollenberger_llm4vv_2024}, an evaluation approach that is explainable, scalable, and very aligned  with human preferences~\cite{zheng_judging_2023}. In this approach, LLMs are assessed for their ability to evaluate and interpret certain outputs. For example, specific frameworks have been developed to test LLMs' capabilities in code evaluation and comprehension~\cite{zhao_codejudge-eval_2025}.

\section{A modular framework for LLM-based code generation evaluation}
\label{sec:framework}

\begin{figure}[t]
    \centering
    \includegraphics[width=\linewidth]{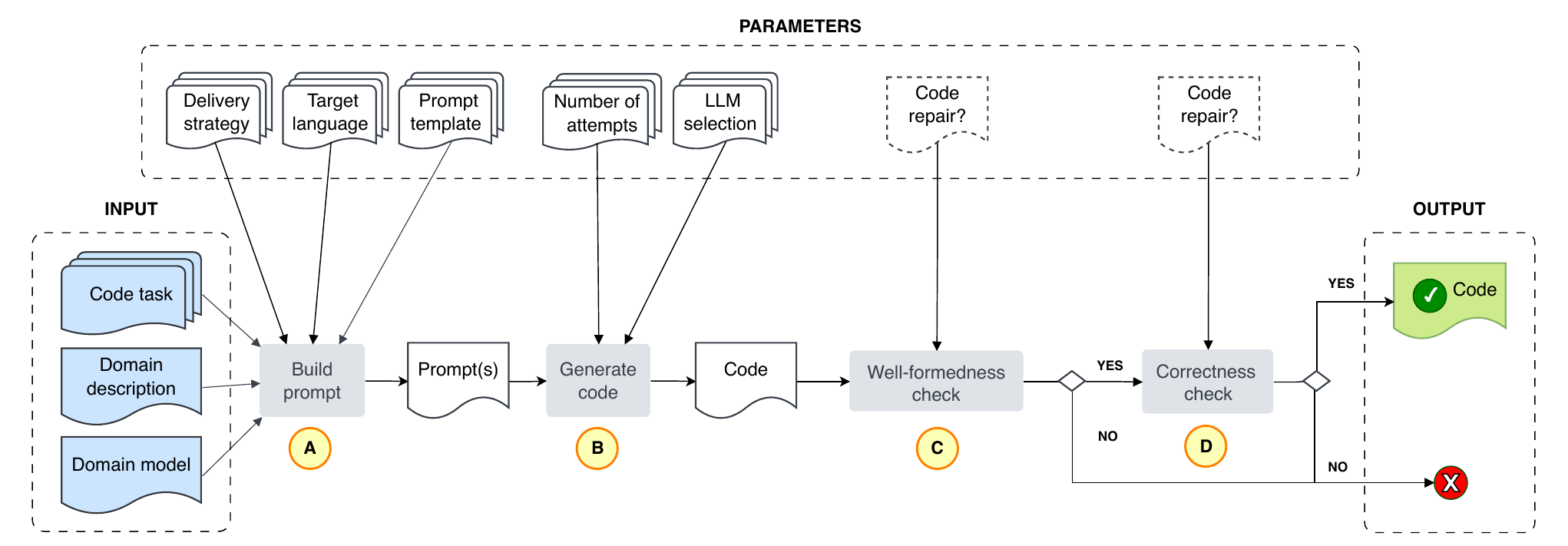}
    \caption{Graphical representation of the framework as a flow diagram\ins{, including labeled components.}}
    \label{fig:flow-diagram}
\end{figure}


This section presents the architecture and components of our proposed framework. \chg{The}{This} framework \chg{first supports}{supports the customization of a code generation process, its execution using the selected prompts and LLMs and} \del{the LLM-based code generation and later }the evaluation of the generated code. 

Figure~\ref{fig:flow-diagram} presents the functional flow, artifacts and tools of our proposed framework. \ins{In all the Figures in this Section,} gray boxes represent an action, orange boxes represent software artifacts, and white shapes represent data \ins{and parameters}.

The inputs to our framework are: (1) the \emph{code task}, which is a natural language description of the specification(s) or query(ies) to be implemented; (2) the \emph{domain description}, using natural language to characterize the relevant features of the domain where the constraint will be evaluated; and (3) the \emph{domain model}, providing a formal description of the domain using a modeling notation. \ins{This framework is very flexible in terms of the inputs to the code generation process: it can operate using only the domain description, only the domain model, or both. This flexibility supports different code generation scenarios.}

Our framework also considers diverse prompting strategies, prompt augmentation techniques and evaluation metrics, offering an automatic evaluation of the quality of their outputs. \ins{These \emph{parameters} can be specified by the user before generating code, allowing the evaluation and comparison of multiple customized versions of the code generation process.}

\ins{This framework is organized around four main stages: building the prompt or prompts used to generate code (block A in Figure~\ref{fig:flow-diagram}, described in Section~\ref{subsec:prompting-and-augmentation}); providing the prompts to the LLM and extracting the generated code (block B in  Figure~\ref{fig:flow-diagram}, described in Section~\ref{sec:code-gen-stage}); checking the well-formedness of the generated code (block C in  Figure~\ref{fig:flow-diagram}, described in Section~\ref{sec:well-formedness-stage}); and, for code snippets that are well-formed, checking their correctness (block D in  Figure~\ref{fig:flow-diagram}, described in Section~\ref{sec:correctness-eval-stage}).}

\ins{In the following subsections, we describe the main stages of the framework in the order in which they are applied.}

\subsection{\chg{Prompting and augmentation}{Building code generation prompts}}
\label{subsec:prompting-and-augmentation}

Given the inputs, our framework builds different prompts following a set of predefined prompt templates. Each prompt template exploits different prompting strategies with the goal to guide the LLM towards the generation of  an appropriate output. 

We define prompting templates that, in order to provide context and guidance to the LLM, operate on two key and orthogonal dimensions:
\begin{enumerate}
    \item \emph{Augmentation}: This dimension leverages Chain of Reasoning (CoR) strategies, which encourage iterative prompting, helping the model achieve a deeper understanding of the task at hand and improving the quality of the generated code.
    \item \emph{Task-Oriented Prompting}: This dimension focuses on designing prompts that accommodate multiple tasks within the same domain, ensuring clarity and precision in guiding the LLM.
\end{enumerate}



The basic prompt construct in our system is a single instruction. Considering that our goal is to evaluate the generation of code for constraints / invariant specifications in a DSL\footnote{Note that our framework would also work for GPLs.}, this construct is structured as: ``\texttt{Code the following constraint
in \{LANGUAGE\}:}'', followed by the coding task provided as input (\emph{e.g.}, ``the number of employees of the company must be less than or equal to 50''). Then, this prompt construct is systematically augmented with additional information to enhance its effectiveness.

Given the characteristics of the code to be generated, our framework incorporates specific elements as part of this augmentation process. These elements, supplied as input to the framework, may include a detailed description of the domain and the domain model itself. The domain model can be formalized in a DSL such as PlantUML or described in natural language. Each template used in the system is designed to evaluate how different combinations of augmentations impact the overall performance of the prompt.

Our framework also supports the option in which the LLM is asked to generate information to enrich the prompt in two different ways: (1) given the inputs, we ask the LLM to explain that domain model; the explanation generated is provided in a subsequent prompt where we request the LLM to solve the coding task; (2) given the inputs, we ask the LLM to generate a domain model either in natural language or following the syntax of a specific DSL (\emph{e.g.}, PlantUML). \ins{Then,} that output is used to enrich the following prompt where we request the coding task to be solved.

\begin{table}[h!]
\centering
\caption{Prompting templates.
}
\small
\begin{tabular}{@{}lccccc@{}}
\toprule
\textbf{PT} & \textbf{Domain Description} & \textbf{Domain Model} & \textbf{Explain DM} & \textbf{Code DM} & \textbf{Prompts} \\ 
\midrule
PT1 & \ding{51} & & & & 1\\
PT2 & & Formal& & & 1\\
PT3 & & NL& & & 1\\
PT4 & & Formal & \ding{51} & & 2\\
PT5 & & NL & \ding{51} & & 2\\
PT6 & \ding{51} & & & NL & 2\\
PT7 & \ding{51} & & & Formal & 2\\ 
PT8 & \ding{51} & Formal & & & 1\\
PT9 & \ding{51} & NL & & & 1\\

\bottomrule
\end{tabular}
\label{tab:promptingtemplates}
\end{table}

Table~\ref{tab:promptingtemplates} illustrates the combination of these augmentations and the nine prompting templates (PT) \chg{that our framework includes}{included in our framework}. The prompts are constructed by concatenating the augmentation components outlined in the table in JSON format. For instance, considering the use case of a DSL for \chg{invariants specification}{the specification of invariants, we would supply the following file to the LLM}:

{\scriptsize
\begin{verbatim}
[
    {
        "role": "user",
        "content": "Consider the context description of the domain: {DOMAIN DESCRIPTION}"
    },
    {
        "role": "user",
        "content": "Code the following constraint in {LANGUAGE}:
        {CODING TASK}"
    }
]
\end{verbatim}
}


The input to our framework may \chg{not always consist of a single coding task: it can also involve multiple tasks,}{also involve several closely related coding tasks}, such as generating \chg{descriptions}{code} for several invariants or constraints \ins{in the same domain}. The second dimension (Task-Oriented Prompting) focuses on \chg{the method of presenting tasks}{how tasks are presented to the LLM} when dealing with multiple \ins{related} tasks within \chg{a single}{the same} domain. \ins{A collection of} tasks can be delivered \ins{to the LLM} in the following ways:

\begin{enumerate}
    \item OP1: Batch delivery. All tasks are presented in the same prompt at once as a list. The expected LLM behavior is that it generates an output that contains a list with the result for each task.
    \item OP2: Chained sequential delivery. Tasks are presented one by one, and the outputs of earlier tasks are incrementally added to the context of subsequent tasks. Note that this requires multiple prompts. For example, the third coding task will \chg{see}{receive} the output of the first and second tasks \ins{as part of its input prompt}, but not the results from the fourth task.
    \item OP3: Isolated delivery. Each task is prompted independently without knowledge of other tasks, focusing on a single, isolated response per task. The output consist of a single coding sample.
\end{enumerate}

\del{Since LLMs are triggered by prompts, we will start the presentation of our framework by focusing on how to build the prompts. Later, we introduce the architecture of our framework and the different components, which are illustrated in Figure~\ref{fig:flow-diagram}.}





%

\subsection{Code generation and extraction}
\label{sec:code-gen-stage}
In this stage, the input coding tasks that are provided are used to build different prompts using the prompting templates described \chg{above (cf. gray box \emph{Build prompt})}{in the previous Section}. These are used to prompt the LLM, producing an output that consists of usually some natural language (NL) description of the code plus the code requested. 

\ins{The user may also request multiple attempts of code generation for the same task. Given that LLMs are not deterministic, each attempt will produce similar code snippets with slight variations among them. Using multiple attempts increases the cost of the code generation process ($k$ attempts means $k$ calls to the LLM), but it also increases the likelihood that one of the generated versions is well-formed and correct (see Section~\ref{sec:code-repair}).}

\chg{This}{The LLM's} output is processed to extract the generated code\del{(cf. gray box \emph{Extract})}. If the prompt contained multiple coding tasks, the response is expected to contain an output for each task hence an additional splitting step \del{(cf. gray box \emph{Split})} is required to isolate \ins{and extract} individual code \chg{samples}{snippets}. 

In our implementation of the framework, this stage is designed to be highly modular. The inputs (\emph{i.e.}, datasets consisting on code tasks, domain description(s) and domain model(s)) are independent of the framework. 
The LLM component used in the framework can be easily replaced, allowing for the code generation evaluation of various LLMs (\emph{e.g.}, GPT-4, Llama, custom fine-tuned models) without altering the overall framework. Similarly, the prompting templates are designed to be extensible, supporting diverse prompting strategies and future extensions.

\subsection{Well-formedness evaluation} 
\label{sec:well-formedness-stage}

\ins{Once we have generated code for the target coding task}, the first step is to verify the well-formedness of the generated code. \ins{A code snippet is considered well-formed when it follows the syntactic rules of the language it is generated.}
This \ins{check} is \chg{done}{performed} using a well-formedness \chg{assertion tool}{validator, such as} a parser or compiler of the language.

Our implementation enables the use of different parsers, compilers, or validation tools depending on the target \chg{DSL (or alternatively GPL)}{language}. This flexibility also applies to the framework capability to integrate different target languages by plugging in the appropriate well-formedness tools. \ins{Depending on the tool used to perform the well-formedness check, it is possible to be assessing the syntactic validity but also semantic correctness at the same time.}

\begin{figure}[htbp]
    \centering
    \includegraphics[width=\linewidth]{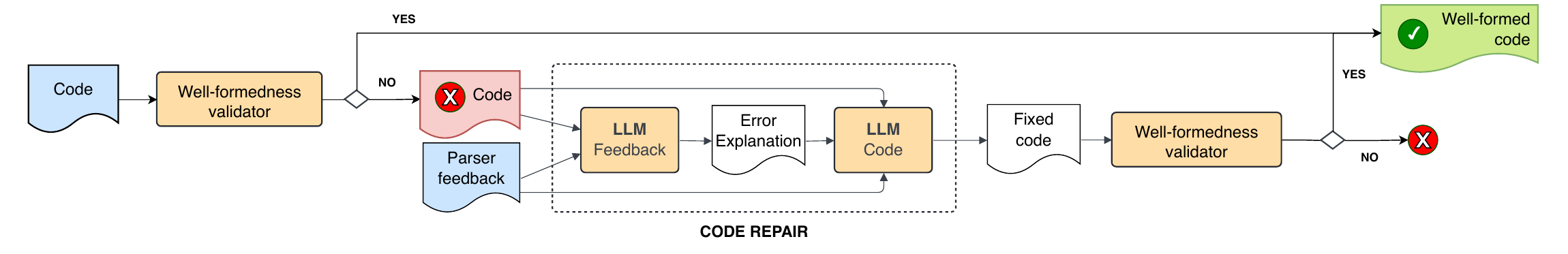}
    \caption{\chg{Well-formedness code repair process diagram.}{Well-formedness checking process diagram (component C in Figure~\ref{fig:flow-diagram}), detailing the well-formedness validation 
    and the well-formedness code repair flow.}} 
    \label{fig:wf-code-repair}
\end{figure}

Code \chg{samples}{snippets} that fail the well-formedness check undergo a \chg{1-pass}{single-pass} code repair process\footnote{A \ins{single-pass or} 1-pass code repair typically refers to a code fixing process that involves making corrections or improvements to source code in a single pass -- that is, reviewing and editing the code only once.}. The error message generated by the well-formedness assertion tool together with the faulty code, is used as input to the code repair process.
Figure~\ref{fig:wf-code-repair} illustrates the process, which \chg{uses}{leverages} Chain-of-Thought \ins{(CoT) reasoning}. \chg{It consists of providing}{First, it provides} the faulty code and the \ins{error message} to an LLM. \ins{Using this information,} the LLM generates an explanation of the error. Then, \chg{ and} this \ins{explanation} is used \ins{by a second LLM}  to generate a new version of the code\ins{, where} \del{(}hopefully \del{in which}the error has been fixed\del{)}. Note that, given the use of CoT, in Figure~\ref{fig:wf-code-repair}, the orange boxes represent two distinct LLMs. However, this \ins{process} could \ins{also} be \chg{done}{performed} using a single \chg{one}LLM and \del{with only}a single prompt\del{, too}. 
Finally, the repaired code undergoes a well-formedness evaluation, as outlined in the previous step. 

\ins{As previously mentioned, we are only performing a single round (or single-pass) of code repair: if the well-formedness validation fails after repair, the coding task is considered failed, without additional rounds of code repair. This is why there is no loop in Figures~\ref{fig:flow-diagram} and \ref{fig:wf-code-repair} if the final well-formedness check fails.} 


\subsection{Correctness evaluation}
\label{sec:correctness-eval-stage}

If a \chg{piece of code}{code snippet} is well-formed, it can proceed to \chg{this}{the following evaluation} stage: the correctness check. \ins{A} \chg{C}{c}ode \ins{snippet} is considered correct when it is executable (\ins{that is}, it is syntactically and semantically correct) and complies with the natural language specification for \chg{which it was created} the target coding task.

\chg{Similar}{Similarly} to the well-formedness stage, the \chg{Correctness Evaluation Tool}{correctness validator} is designed as a plug-and-play component. This allows for the integration of various verification methods, such as unit testing frameworks (\emph{e.g.}, AUnit for Alloy), formal verification tools, or custom correctness checkers, depending on the requirements of the specific language and code task.

\begin{figure}[htbp]
    \centering
    \includegraphics[width=1\linewidth]{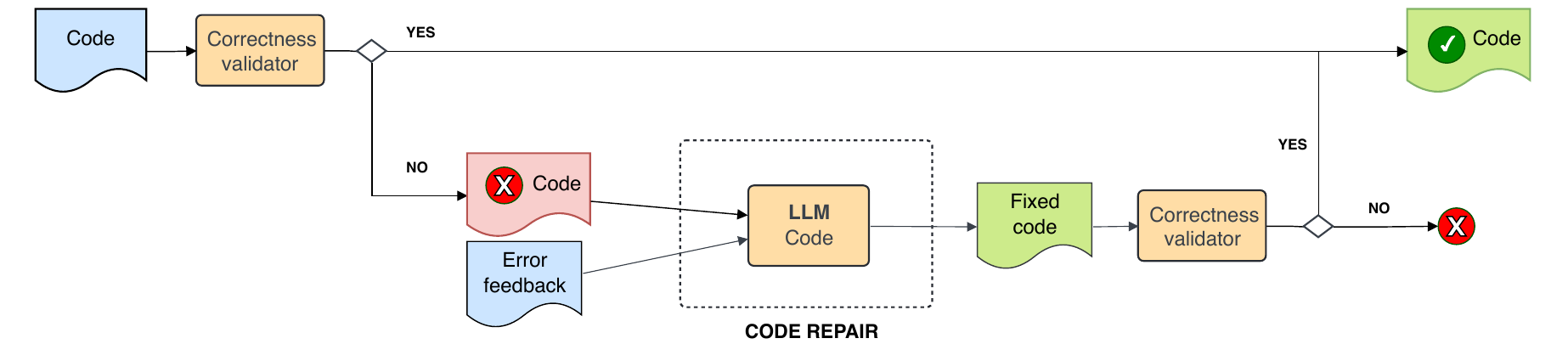}
    \caption{\chg{Correctness code repair process diagram.}{Correctness checking process diagram (component D in Figure~\ref{fig:flow-diagram}), detailing the flow between the correctness validator (see Figure~\ref{fig:llm_as_a_judge}) and the correctness code repair process.}}
    \label{fig:wf-code-repair-correctness}
\end{figure}

Code \chg{samples}{snippets} that fail the correctness check undergo a \chg{1-pass}{single-pass} code repair process, which implies that code repair occurs only once. The process is shown in Figure~\ref{fig:wf-code-repair-correctness}: the error messages / feedback from the correctness \chg{evaluation tool}validator, along with the faulty code is provided as input to the LLM. The LLM \ins{then} generates a fixed version of the code, which is sent back for correctness check. \ins{Similarly to what happens in well-formedness repair, we only perform a single round of code repair, so the process ends if the final correctness check fails.}

\ins{While the notion of correctness we are using is binary (either a code snippet is correct or not), when we are considering multiple code generation attempts there are different ways to measure the success of code generation. In the following, we describe the two metrics that we use in the experiments in this paper:} pass@k and accuracy.


\paragraph{Pass@k}\chg{When given a specification, the code is generated more than once, }{When using multiple code generation attempts,} pass@k evaluates the probability that at least one of the \textit{k} generated samples is correct. Formula \ref{eq:pass@k} shows how pass@k is calculated, considering that \textit{n} represents the total number of input examples, \textit{c} represents the number of correct samples, \textit{k} is them number of samples selected, and \textit{C} is a combination function. By computing the probability that all selected samples are incorrect and subtracting that from \textit{1}, we obtain the probability that at least one is correct. 

\begin{equation}
\label{eq:pass@k}
\text{pass@k} = 1 - \frac{C(n - c, k)}{C(n,k)}
\end{equation}

\paragraph{Accuracy} \del{When given a specification, the code is generated more than once, the} Accuracy measures the number of times \ins{in which} the generated code \chg{exhibits correctness}{is correct} in the first attempt. The formula of accuracy is shown in the equation \ref{eq:accuracy}. 

Note that both pass@k and accuracy are related as accuracy is equal to pass@1, when \textit{k = 1}.

\begin{equation}
\label{eq:accuracy}
    accuracy = \frac{correct~examples}{total~number~of~examples} = pass@1
\end{equation}
\section{Framework instantiation}
\label{sec:instantiation}

To assess the effectiveness of our approach, we have applied it to two DSLs for constraint languages: OCL and Alloy. Our framework supports any language. \del{hence,} For the sake of comparison between low-resource and high-resource languages, we have also applied it to assess one \chg{GLP}{general-purpose language}: Python. In the following, we explain the instantiation of our framework to these languages.

The \ins{well-formedness and correctness validators}\del{evaluation components of our framework (that is, the components \emph{check} and \emph{repair})} are language-specific. In this section, we explain the tools and approaches we have used to evaluate both the well-formedness and correctness of these three languages.

\subsection{Well-formedness evaluation}
\label{subsec:well-formedness_evaluation}
We have implemented two distinct approaches for the evaluation of well-formedness:

\begin{enumerate}
    \item Automatic syntactic validation: 
    This approach employs parser implementations derived from existing grammar repositories, specifically the ANTLR grammars for OCL, Alloy, and Python available at \cite{antlr_project_grammars-v4ocl_2025}, \cite{antlr_project_grammars-v4alloy_2025} and \cite{antlr_project_grammars-v4pythonpython3_13_2025}. These parsers are used to verify the syntactic correctness of the code. \chg{An}{A} piece of code is deemed well-formed if the parser completes without reporting syntax errors. 

    \item Manual validation via tool execution:
    This is complementary to the automatic syntactic validation. In this case, the code is executed within dedicated environments (\textit{USE OCL} for OCL, \textit{Alloy Analyzer} for Alloy) and Python \ins{interpreters}, which \chg{is}{are} used to manually validate the well-formedness of inputs through dynamic compilation and execution. This manual inspection involves reviewing the error messages produced by these tools during execution to determine well-formedness. This manual validation captures errors that may not be captured by syntactic parsing alone because the tool execution also performs a semantic validation.
\end{enumerate}

\subsection{Correctness evaluation}
\begin{figure}[t]
    \centering
    \includegraphics[width=.8\linewidth]{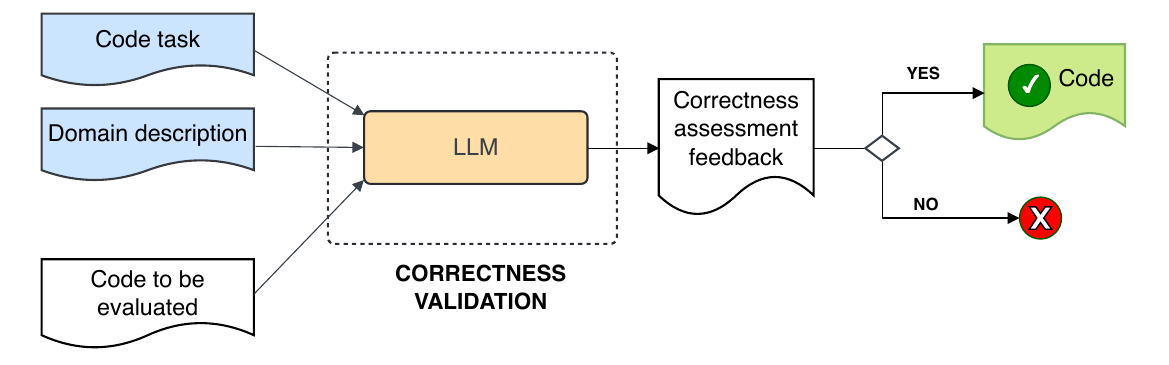}
    \caption{\chg{LLM-as-a-judge approach description.}{Correctness validator. Overview of the validation process via our LLM-as-a-judge framework.}}
    \label{fig:llm_as_a_judge}
\end{figure}

To assess the correctness of the generated code we used the following approaches:
\begin{enumerate}
    \item Automatic evaluation employing an LLM-as-a-Judge\del{ technique}~\cite{gu2025surveyllmasajudge}: This approach leverages a large language model to automatically evaluate whether the generated code meets the expected correctness criteria. We used the same LLM (GPT-4o-mini) to assess every correctness evaluation for every target language, to obtain comparable results.

    We have implemented LLM-as-a-Judge as described in Figure~\ref{fig:llm_as_a_judge}. The LLM evaluates the correctness of a generated piece of code (\emph{code to be evaluated}) using the inputs \emph{coding task specification in natural language} and \emph{domain description}. If the generated code \chg{resulted}is incorrect, the correctness assessment feedback is sent to the correctness code repair phase. \del{Otherwise, the constraint is considered checked.}
    
    \item Manual evaluation based on specification fulfilment: In this approach, a human evaluator manually assesses whether the generated code satisfies the intended requirements, either by verifying adherence to the input specification or by confirming that the code correctly implements the specified constraint.
\end{enumerate}

\subsection{Practical considerations}

\paragraph{Language-specific guidelines} Since this paper puts the focus on constraints / invariants, it is worth mentioning that some languages can describe a given constraint in many alternative ways. In order to facilitate the analysis of the generated code (\emph{e.g.}, to make it easier to locate the generated constraints inside the fragment of generated code), we provide language-specific instructions to guide how constraints should be described:

\begin{itemize}
\item Python: \chg{We request the LLM to generate a function for each constraint. This function will return a boolean value stating whether the constraint is satisfied or not.}{We instruct the LLM to generate one function per constraint, returning a boolean value indicating whether the constraint is satisfied. This restriction is necessary because Python is a general-purpose language. As a result, without any guidance, the LLM tends to generate entire program fragments implementing the constraint rather than a concise constraint checker, resulting in heterogeneous and non-comparable outputs.} 
\item Alloy: We request the LLM to generate each constraint as a single and unique \texttt{fact}. \ins{Alloy also offers multiple constructs capable of expressing constraints (\emph{e.g.}, facts, predicates and assertions), and our initial experiments showed a diversity of structural variations. Standardizing constraints as facts ensures consistency across examples and aligns the Alloy representation with those produced in Python and OCL.} 
\item OCL: No language-specific guideline is needed. 
\ins{OCL is designed explicitly as a constraint specification language, and invariants are expressed using a dedicated grammatical construct (\texttt{context} \textit{<Classifier>} \texttt{inv} \textit{<Name>: <Expression>}). Because the language has a canonical form for invariants, the LLM consistently generated constraints in the expected form without additional instructions. Consequently, constraints in OCL are naturally uniform and require no further guidance.} 
\end{itemize}


\paragraph{Large Language Models}
The framework is designed to be model-agnostic and does not depend on a specific LLM provider or API. It supports a modular architecture that allows for the integration of multiple language models through adapter components, enabling straightforward substitution or comparison of different LLMs. In this work, we have implemented and tested the framework using four models: DeepSeek Coder 6.7B, Llama 3.1, GPT-4o, and GPT-4o-mini. \ins{When generating multiple attempts, in order to ensure non-deterministic behavior, we have to set the \emph{temperature} parameter of the LLM to a value that enables some degree of flexibility in the outputs. For the experiments in this paper, we have used a temperature of 0.3, which introduces a mild stochasticity in the generated code, while avoiding outputs that are repetitive and less representative of the model's behavior in practice.}

\paragraph{Context window}
LLMs are constrained by the size of their context windows. For instance, GPT-4o supports up to 128,000 tokens, which is substantially larger than the context window available in smaller models, such as DeepSeek Coder 6.7B.
If a prompting strategy involves a domain model of significant size, it is very likely that the resulting prompt will exceed the context window of the evaluated LLMs. One commonly used technique to address such scenarios is retrieval-augmented generation (RAG) \cite{pan_generative_2024}. However, this framework does not currently implement or rely on RAG.

\paragraph{Replicability package} Our implementation of the framework, together with all the experiments reported in the following section, is available on our Github repository~\cite{repo}.
\section{Validation}
\label{sec:case-study}
  

In this section, we describe the experiments performed to assess the feasibility of our framework when studying the effectiveness of LLMs for code generation tasks.
In Section~\ref{sec:manual-automatic}, we present an initial experiment to assess the performance of our code evaluator.
In Section~\ref{sec:datasets}, we describe the datasets  used in our experiments. Using the instantiation and implementation of our framework and these datasets, we study the different factors that may impact the quality of the generated code. These are:
    target language (Section~\ref{sec:res-language});
    \chg{used}{choice of} LLM (Section~\ref{sec:llm});
    prompt delivery approach (Section~\ref{sec:prompting});
    task delivery approach (Section~\ref{sec:task-delivery});
    and code repair (Section~\ref{sec:code-repair}).
We conclude by \ins{discussing the results and key takeaways in Section~\ref{sec:discussionAndTakeaways}} and  discussing threats to validity in Section~\ref{sec:threats}.



%
%

\subsection{Manual vs automatic evaluation
}
\label{sec:manual-automatic}


Our framework compares multiple strategies to generate code, considering different LLMs, prompt delivery strategies, use of code repair, \ldots However, the sheer number of combinations quickly makes it impossible to use human evaluation to compare all the generated code snippets. Thus, it is necessary to use techniques for automatic code evaluation. \ins{As discussed in Section~\ref{sec:rel-eval}, this evaluation should not consist of measuring the similarity with respect to a gold-standard reference solution, as this does not properly consider valid alternative implementations of the same constraint~\cite{DBLP:journals/tse/CrupiTVMPB25,DBLP:journals/tosem/HuCWXLZ22}. Instead, we should focus on assessing how faithfully the invariant implements the intended constraint.} 

In this section, we discuss the approaches that we have used to automatically assess the well-formedness and correctness of the generated code. Given that the evaluation step is critical for our experiments, we will measure the quality of this automatic evaluation, comparing it to a \chg{human}{manual} evaluation \ins{performed by a human expert}. 

\paragraph{\ins{Manual evaluation}} 

\ins{The manual evaluation has covered 1,512 code snippets: 56 coding tasks, generated with 9 different prompt templates for 3 target languages: OCL, Alloy and Python. The protocol used for the manual evaluation has been the following. One author (a PhD candidate with experience in model-driven engineering and AI) has inspected the code snippets, identifying representative examples and common problems. These examples have been discussed among all three co-authors to identify common evaluation criteria, some of which are language-specific. For example, the use of non-existing libraries in OCL in the generated code (\emph{e.g.}, a \texttt{Date} type) has been considered a correctness problem; similarly, in Alloy any incorrect usage of multiplicities (such as \texttt{one} or \texttt{lone}) or the subset operator \texttt{in} has been considered a correctness problem rather than a well-formedness issue. Using this set of criteria, the manual evaluation of the examples was carried out by the same author that selected the initial set of samples.} 

\ins{Well-formedness has been checked in two stages. First, an ANTLR parser for the target language has been used to identify whether the code is syntactically correct. Then, all code snippets that are syntactically correct have been validated on an analysis or execution environment for the target language: the UML-based Specification Environment (USE) for OCL, the Alloy Analyzer for Alloy and a Python 3.9 interpreter for Python. These tools can identify semantic errors such as type errors or naming problems. If the corresponding tool is able to read the code snippet without errors, the code is deemed well-formed.}

\ins{Correctness has been checked only on well-formed code snippets. This step has been performed manually, with the assistance of the previously mentioned analysis and execution tools. For instance, in some of the more complex constraints, we have populated the domain models to inspect the effect of the generated constraints.}



\paragraph{\ins{Automatic evaluation}} \del{In order to measure the quality of the automatic evaluation,} \chg{we}{We} have assessed the performance of GPT-4o-mini\ins{, DeepSeek3.1, o3, and Llama-3.3-70B} as the LLM judge\ins{s} by comparing its \chg{success}{validation results} with the \chg{judge of a human expert}{results of the manual evaluation}. 
\del{We have used 1,512 pieces of generated code for Alloy, OCL and Python, which have been manually reviewed and assessed by one of the authors of this paper.}
%
%
To evaluate how well the generated code aligns with our manual evaluation, we frame the problem as a binary classification problem, \emph{i.e.}, the code is \chg{or is not correct}{either correct or incorrect}.

\paragraph{\ins{Measuring the quality of an evaluation}} Although accuracy is usually used to study the quality of AI models, it requires a balanced dataset. Otherwise, it can give a false sense of model performance by favoring the majority class. Since this is not the case in our experiments (cf. Section \ref{sec:datasets}), 
we use precision and recall as our primary evaluation metrics. 

\chg{The precision is calculated as the ratio given by dividing the number of fragments of code marked as correct by both human and automatic evaluation by the number of fragments of code marked correct by the automatic evaluation. Equation~\ref{eq:precision} presents the formula.}{Precision is the proportion of true positives among all positive predictions. Equation~\ref{eq:precision} presents the formula, in which \emph{TP} (true positives) are code fragments marked as correct by both the manual evaluation and the automatic check. \emph{FP} (false positives) are code fragments marked as correct by the automatic check but incorrect according to the manual evaluation.}

\begin{equation}
\label{eq:precision}
\mathrm{Precision} = \frac{\text{| code considered correct by both (TP) |} }{\text{| code considered correct by automatic eval. (TP + FP) |}}
\end{equation}

\chg{Recall is calculated as the ratio given by the ratio given}{Recall is calculated as the ratio given by dividing the number of fragments of code marked as correct} by both human and automatic evaluation \ins{(TP)} by the number of fragments of code marked correct by the human \ins{(TP + FN)}. Equation~\ref{eq:recall} presents the formula\ins{, where FN (false negatives) are code fragments marked as incorrect by the automatic evaluation but marked correct by the manual evaluation}.

\begin{equation}
\label{eq:recall}
    Recall = \frac{\text{| code marked as correct by both (TP) |}}{\text{| code considered correct by human (TP + FN) |}}
\end{equation}

\paragraph{\ins{Discussion of the results}}  Table~\ref{tab:evaluation_metrics} 
\chg{compares the results of GPT-4o-mini with those of a human expert in terms of precision and recall. We can see that precision and recall for well-formedness and correctness are close, which indicates that the evaluation neither misses many correct code nor overestimates correctness.}{ compares the results of 4 LLM-based judges: GPT-4o-mini, DeepSeek, o3, and Llama-3.3. Overall, all models achieve comparable performance, with precision and recall typically ranging between 0.7 and 0.85. However, notable differences emerge across languages. In Python, o3 exhibits a larger gap between precision and recall, indicating a tendency to overestimate correctness relative to GPT-4o-mini, which shows a more balanced trade-off. In contrast, Llama-3.3 demonstrates both higher precision and recall for Python, resulting in the most balanced performance among the evaluated judges for this language. For Alloy and OCL, the differences between judges are smaller, though DeepSeek consistently attains higher recall, suggesting a more permissive evaluation strategy.}

\ins{Moreover, these results show a disagreement around 15-30\% in precision and recall for the best models. It should be noted that in complex AI/ML tasks such as machine translation or code generation, some level of disagreement is expected even among human experts~\cite{DBLP:journals/tacl/FreitagFGRTM21}. For instance, \cite{DBLP:journals/tosem/HuCWXLZ22} reports Pearson correlations between human annotators ranging from 0.7 to 0.9 when rating code commit descriptions, indicating strong but imperfect agreement. Similarly, \cite{zheng_judging_2023} shows that strong LLM judges such as GPT-4 achieve high alignment with human preferences on MT-bench and Chatbot Arena benchmarks, including coding tasks, with 85\% agreement between GPT-4 and humans—exceeding the 81\% agreement observed among humans themselves. These results suggest that discrepancies between LLM-as-a-judge evaluations and human solutions largely reflect the inherent variability of human judgment rather than deficiencies in automated evaluation.}

\ins{Overall, the results indicate that LLM-as-a-judge approaches provide a reliable approximation of human correctness judgments, with no single model uniformly dominating across all languages and metrics. For GPT-4o-mini and Llama-3.3, the precision and recall values are similar. This suggests that automatic evaluation neither substantially overestimates correctness nor misses many correct solutions. That is, errors are spread equally among false positives and negatives,  supporting the robustness of LLM-based judging in this setting.}

It is worth mentioning that\del{,} the concrete values of precision and recall for \del{the} well-formedness are \chg{not closer to 1}{lower} due to the way in which the evaluations have been performed. The automatic evaluation of well-formedness implemented in the framework relies solely on a parser and checks only syntactic correctness, whereas the manual evaluation was performed by executing the code, therefore it considers both syntax and semantics (\emph{i.e.}, conformance to the domain model). As such, the manual evaluation is stricter, which creates some discrepancies\del{among}. 

\begin{table}
\centering
\caption{Performance of automatic code evaluation for determining well-formedness and correctness across different LLMs}
\label{tab:evaluation_metrics}
\scriptsize \smallskip
\begin{tabular}{lrrcccccccc}
\toprule
& \multicolumn{2}{c}{\textbf{Well-formedness}} 
& \multicolumn{8}{c}{\textbf{Correctness}} \\
\cmidrule(r){2-3}
\cmidrule(l){4-11}
\textbf{Language} 
& \textbf{Precision} 
& \textbf{Recall} 
& \multicolumn{2}{c}{\textbf{GPT-4o-mini}}
& \multicolumn{2}{c}{\textbf{DeepSeek}}
& \multicolumn{2}{c}{\textbf{o3}}
& \multicolumn{2}{c}{\textbf{Llama-3.3}} \\
\cmidrule(lr){4-5}
\cmidrule(lr){6-7}
\cmidrule(lr){8-9}
\cmidrule(lr){10-11}
& & 
& \textbf{Prec.} & \textbf{Rec.}
& \textbf{Prec.} & \textbf{Rec.}
& \textbf{Prec.} & \textbf{Rec.}
& \textbf{Prec.} & \textbf{Rec.} \\
\midrule
Alloy 
& 0.574 & 0.504 
& 0.763 & 0.798 
& 0.814 & 0.843 
& 0.809 & 0.819 
& 0.754 & 0.762 \\
OCL 
& 0.786 & 0.853 
& 0.696 & 0.688 
& 0.664 & 0.799 
& 0.706 & 0.826 
& 0.717 & 0.731 \\
Python 
& 1.000 & 1.000 
& 0.770 & 0.834 
& 0.571 & 0.857 
& 0.611 & 0.849 
& 0.774 & 0.859 \\
\bottomrule
\end{tabular}
\end{table}

\begin{figure}
    \centering
    \includegraphics[width=\linewidth]{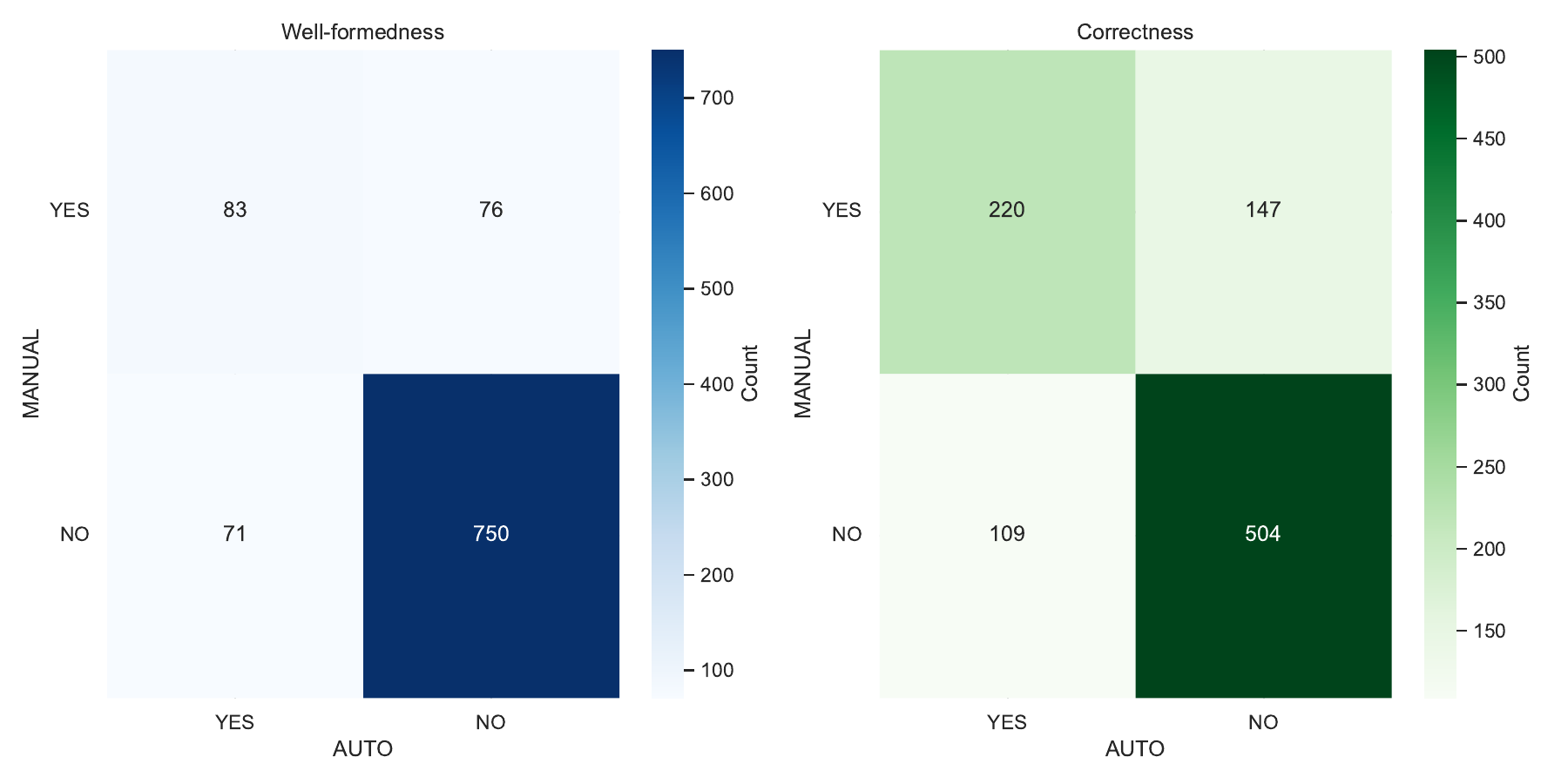}
    \caption{Confusion matrices for well-formedness and correctness \ins{using GPT-4o-mini as the LLM judge}.}
    \label{fig:confusion-matrices}
\end{figure}

To provide further details about automatic evaluation, Figure~\ref{fig:confusion-matrices} presents two confusion matrices. The left one illustrates the alignment of our parser-based automatic evaluation of well-formedness with a human expert evaluation. Meanwhile, the second matrix (right) illustrates the alignment between manual evaluation and the LLM-as-a-judge assessment in the case of correctness.\footnote{For the sake of conciseness, we only present the confusion table for the experiments with GPT-4o-mini.}
In both matrices, rows represent the manual evaluation outcomes, while columns indicate the \ins{result of} our automatic check of well-formedness or correctness. The top-left cell (true positives) counts cases where both manual evaluation and the automatic check agree that the criterion is met (it passes the check). The top-right cell (false negatives) records instances where the manual evaluation decides that the criterion is met, but the automatic check disagrees. The bottom-left cell (false positives) represents cases where the manual evaluation indicates that the criterion is not met, but the automatic check suggests otherwise. Lastly, the bottom-right cell (true negatives) captures cases where both manual evaluation and the automatic check agree that the criterion is not met (it does not pass the check).

True positives and negatives dominate these confusion matrices, reflecting the high levels of agreement that were previously discussed. In conclusion, even if the proposed automated evaluation strategy is not perfect, it achieves a reasonable degree of precision. Thanks to it, we have been able to evaluate \chg{98,397}{thousands} of code generation tasks using several combinations of code generation strategies, which would have been impossible to do manually. \ins{Section \ref{sec:results} provides a detailed description of the experiments performed and the variables being studied.} 

\subsection{Datasets used in the experiments}
\label{sec:datasets}



For the evaluation of the framework we have employed two datasets adapted from prior studies \cite{abukhalaf_pathocl_2024, pan_generative_2024} to cover a wide range of constraint scenarios, ensuring broad applicability. It is worth noting that the dataset in \cite{pan_generative_2024} is not fully used due to problems which are unrelated to our framework itself, such as exceeding the limit of the context window size of the particular LLMs that we have selected for evaluation.

\begin{figure}[htbp]
    \centering
    \includegraphics[width=1\linewidth]{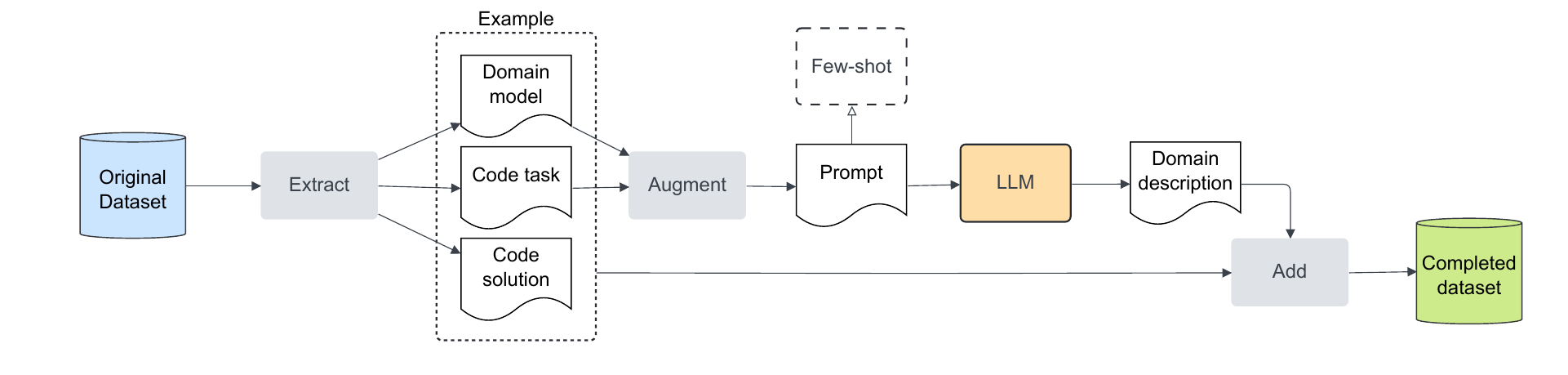}
    \caption{Dataset completion process}
    \label{fig:dataset-completion}
\end{figure}

Unfortunately, all the datasets that fit our purpose and \chg{we}{were} considered \del{using} lack domain descriptions in natural language. To \chg{provide}{enrich} them with the descriptions, we have followed the process illustrated in Figure~\ref{fig:dataset-completion}. From the original dataset, we have extracted the domain model, coding tasks, and code solutions. Then, using an LLM independent of those used for evaluation \ins{(in our case, Google Gemini 2.5 Pro)}, \chg{have we}{we have} synthetically generated domain descriptions. \ins{There were a total of 30 domains (15 per dataset) where the generation of a domain description was necessary. The prompt\footnote{The complete prompt can be found at this link: url{https://github.com/david-xander/code-generation-framework-for-dsls/blob/main/datasets/generate\_domain\_context\_prompt.md}} that requests a domain description includes the PlantUML model in textual format as well as the set of constraints in the model.} 

\del{To do this, for each domain model and code tasks associated to it, we have created a prompt, which has been enriched by including examples of the kind of descriptions that we were aiming for to guide the LLM.}

\ins{The datasets being used were originally designed for the evaluation of generated OCL code. To this end, they provide a reference solution for each of the constraints in OCL. However, they do not provide reference solutions for other languages such as Alloy or Python. As mentioned in Section \ref{sec:manual-automatic}, this not an issue as we will be using LLM-as-a-judge rather than a NLP score that compares generated solution with a gold-standard reference solution.}

\begin{table}[ht]
\centering
\caption{Datasets used in the experiments: domains and features of the class diagram for each domain}
\label{tab:ds_metrics_summary_with_averages}
\scriptsize
\begin{tabular}{lcccccccc}
\toprule
\multirow{3}{*}{\textbf{Dataset}} & \multicolumn{2}{c}{\textbf{Domains}} & \multicolumn{6}{c}{\textbf{Class diagrams}}  \\
\cmidrule(lr){2-3} \cmidrule(lr){4-9} 
 & \textbf{\#} & \textbf{Word count} & \textbf{Classes} & \textbf{Associations} & \multicolumn{2}{c}{\textbf{Word count (average)}} &  \multicolumn{2}{c}{\textbf{Constraints}}  \\ \cmidrule(lr){6-7} \cmidrule(lr){8-9}
  &  &  &  &  & \textbf{ Natural language} & \textbf{Formal} &  \textbf{\#} & \textbf{Word count} \\
\midrule
DS1 & 15 & 169.13 & 253 & 431 & 325.60 & 251.87 & 86 & 17.36 \\
DS2 & 13 & 188.08 & 75 & 27 & 261.31 & 98.92 & 96 & 12.49 \\
\bottomrule
\end{tabular}


\smallskip
\textbf{Word count}: Average number of words used in the textual representation of each category. For example, constraints' word count is the average number of words required to describe the constraints in the dataset. Other columns represent dataset metrics as labeled.

\end{table}

Table~\ref{tab:ds_metrics_summary_with_averages} summarizes the characteristics of the domain models used \ins{in the experiments}: the number of models in each dataset and the total number of classes, associations, and integrity constraints. Moreover, for each of these elements, we report the average length (in words) of its description in textual format. In the case of class diagrams, they can be described using natural language or a formal notation such as PlantUML, so the table provides both word counts. The word count information is relevant, for instance, to compare the number of tokens used by different prompting strategies.

\subsection{Results}
\label{sec:results}

By using our framework and framework instantiation, in the following, we study and describe the different factors that can affect the quality of the generated code using the previously described datasets.

\ins{The detailed breakdown of our experiments is as follows: we evaluate two datasets, including a total of 30 different domain models and 182 constraints. Each experiment was performed using 4 different LLMs, although open source LLMs were unable to perform some tasks due to their limited context window which was insufficient to process the prompt and context information. Each experiment was targeting one of 3 different languages (OCL, Alloy and Python). Experiments were performed with a variety of settings, in order to measure their impact in the well-formedness and correctness of the generated code: 9 different prompting templates, 2 task-delivery methods (batch vs isolated) and 2 code repair strategies (with and without code repair). Moreover, we also measured the effects of increasing the number of attempts in each experiment (pass@k, using $k$=3). As a whole, a total of 98,397 code generation tasks were evaluated as part of these experiments.} 

\subsubsection{Does code generation depend on the target programming language?}
\label{sec:res-language}

\paragraph{Hypothesis} The efficacy of LLM in code generation is influenced by the target programming language.

\paragraph{Experiments} We compare the quality of the code generated for the two DSLs OCL and Alloy to a general-purpose language like Python.

\begin{figure}[ht]
    \centering
    \begin{subfigure}[b]{0.48\linewidth}
        \centering
        \includegraphics[width=\linewidth]{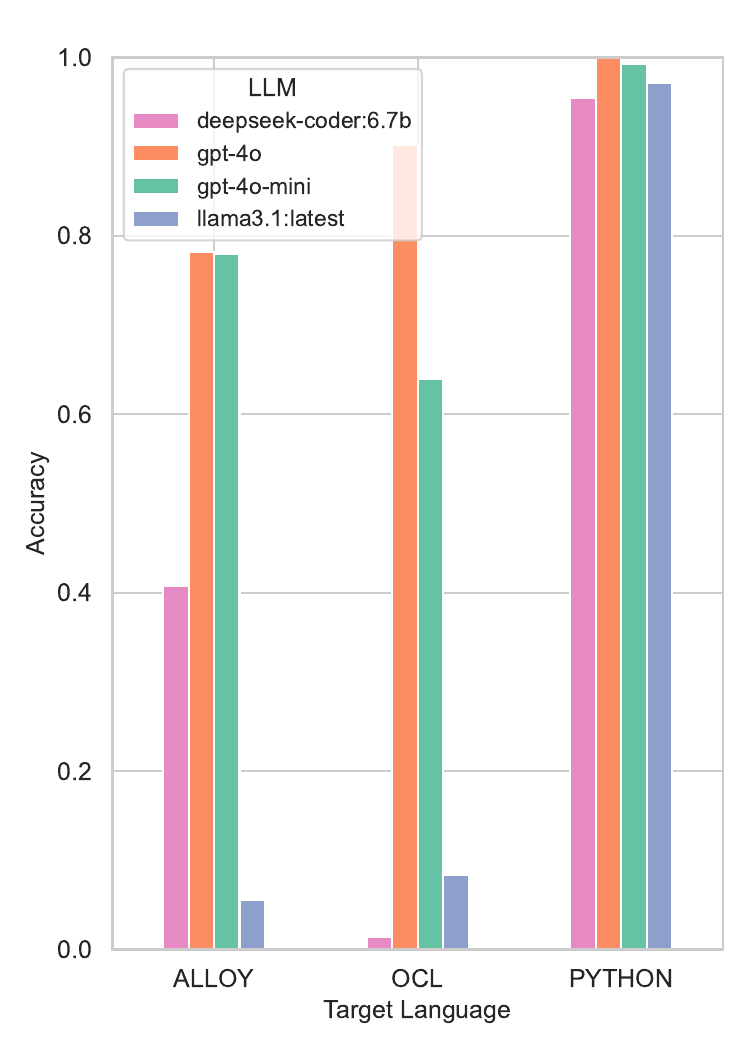}
        \caption{Well-formedness}
    \end{subfigure}
    \hfill
    \begin{subfigure}[b]{0.48\linewidth}
        \centering
        \includegraphics[width=\linewidth]{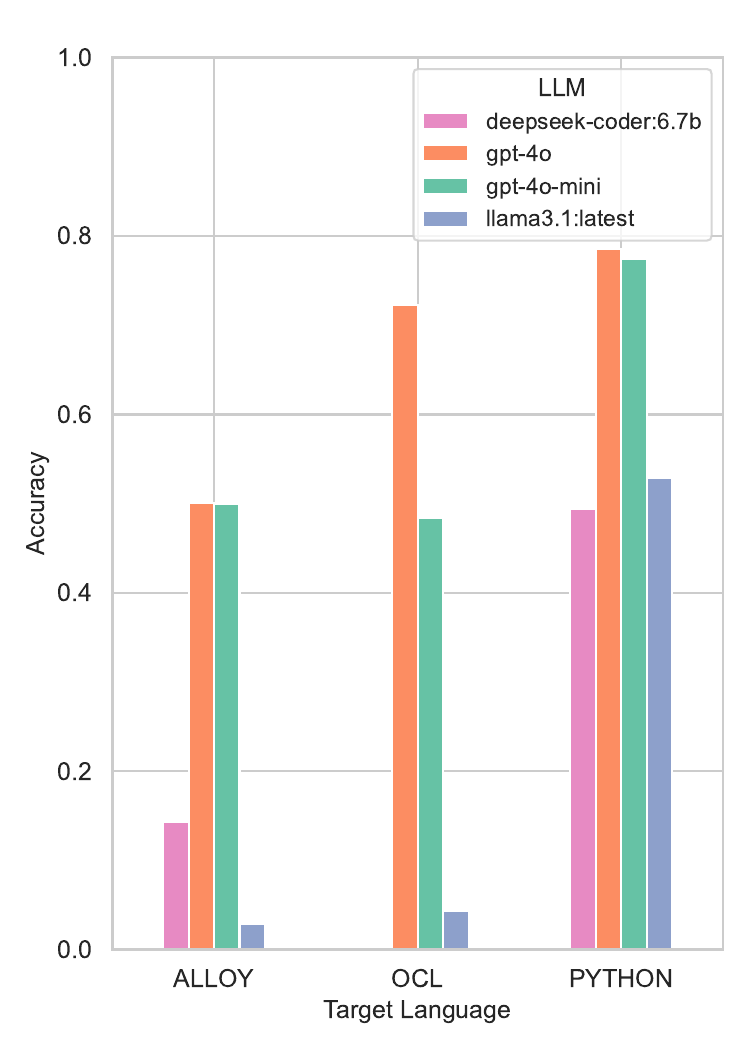}
        \caption{Correctness}
    \end{subfigure}
    \caption{\chg{Well-formedness and correctness for code generation for different LLMs and target languages.}{Well-formedness and correctness of code generation across LLMs and target languages. Accuracy (defined in Section \ref{sec:correctness-eval-stage}) is shown for each LLM–language pair, with four bars per target language.}}
    \label{fig:different-languages-performance-across-datasets}
\end{figure}

\paragraph{Results} Figure~\ref{fig:different-languages-performance-across-datasets} shows the results for code generation targeting OCL, Alloy and Python.

Python achieves the best results both for well-formedness and correctness. Almost 100\% of the Python code generated in the experiments is well-formed, but correctness is only achieved in (at most) 80\% of the experiments. Meanwhile, Alloy and OCL already have problems with well-formedness. \chg{Morever}{Moreover}, even when excluding code that is not well-formed, Alloy and OCL code has more errors \chg{that make it incorrect}{affecting its correctness}.

Python exhibits superior semantic coherence and broader applicability across tasks compared to the DSLs. However, this advantage diminishes for tasks closely aligned with the inherent purposes of the DSLs, \emph{e.g.}, OCL has less of a disadvantage when generating integrity constraints that use set operators and navigations through associations.

Coding tasks focusing on multiplicity (\emph{e.g.}, \texttt{size()} or \texttt{notEmpty()} in OCL and \texttt{one}, \texttt{lone}, \texttt{some} in Alloy) were generally well-handled in all the target languages. Nevertheless, derived attributes and tasks requiring inference from implicit logic pose significant challenges for LLMs.

\paragraph{Discussion} There are noticeable differences in the quality of the generated code across languages.
There are several issues that seem to contribute to the problems being identified:
\begin{itemize}
\item Alloy and OCL include some niche operators that are not common in general-purpose programming languages. An example could be the set and relational operators from Alloy. These operators have very specific typing restrictions and semantics, making them hard to use without having access to examples.
For instance, Alloy's subset operation  \texttt{in} caused many well-formedness problems because LLMs were incapable to use it properly, \emph{e.g.}, comparing sets of different type.
\item Unlike general-purpose programming languages, Alloy and OCL lack a standard library that offers frequent types for managing data types such as dates. As a result, in experiments where dates \chg{were}{are} involved, LLMs often \chg{hallucinated}{hallucinates} functions that could perform date and time operations.
\end{itemize}

\subsubsection{Does the quality of the generated code depend on the LLM being used?}
\label{sec:llm}

\paragraph{Hypothesis} Different LLMs lead to different solutions of different quality.

\paragraph{Experiments} We have compared four different LLMs: two proprietary ones by OpenAI (GPT-4o, GPT-4o-mini) and two open-source LLMs (DeepSeek Coder 6.7b and Llama 3.1). 

\paragraph{Results and discussion} Figure~\ref{fig:different-languages-performance-across-datasets} illustrates the relative performance of the four different LLMs for code generation tasks.

Looking into the results, GPT-4o is the best when considering both well-formedness and code quality. These results are closely followed by GPT-4o-mini, which only drops in performance when considering OCL code. This seems to indicate that GPT-4o has been trained with OCL snippets (or, at least, some OCL-like textual notation) while GPT-4o-mini has not had this information available in its training dataset.

In contrast, DeepSeek and Llama 3.1 exhibit much worse results, having serious problems even generating well-formed code. DeepSeek has an edge when generating Alloy code, but even then the quality of the results is poor: 40\% of well-formed code snippets for Alloy and, within those, less than 15\% of the correct ones. This makes open source LLMs unusable in practice for code generation tasks targeting constraint-focused DSLs. For this reason, for the remainder of the experiments we will be using GPT-4o and GPT-4o-mini as the LLMs of choice.


\subsubsection{\chg{What}{Which} prompting technique performs best?}
\label{sec:prompting}

\paragraph{Hypothesis} As explained in Section \ref{sec:framework}, our framework provides different prompt templates to describe the code generation task and provide the relevant contextual information. Our hypothesis is that the prompting technique might have an impact on the quality of the result.

\paragraph{Experiments} Our framework by default uses all the prompting templates described in Table~\ref{tab:promptingtemplates}. We have used them for the OpenAI GPT models (GPT-4o and GPT-4o-mini) and for each target language (Alloy, OCL and Python).

\begin{figure}[t]
    \centering
    \begin{subfigure}[b]{0.32\linewidth}
        \centering
        \includegraphics[width=\linewidth]{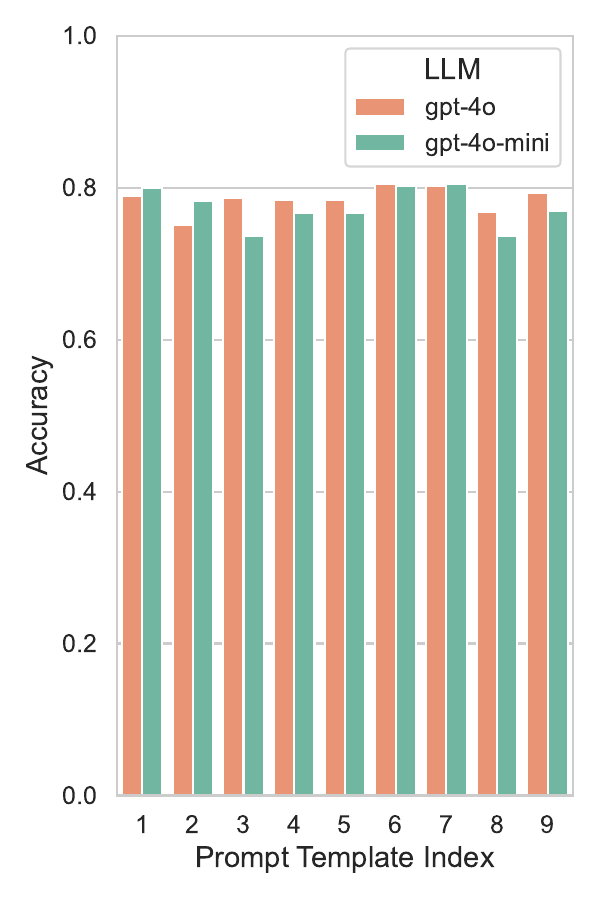}
        \caption{Python.}
    \end{subfigure}
    \hfill
    \begin{subfigure}[b]{0.32\linewidth}
        \centering
        \includegraphics[width=\linewidth]{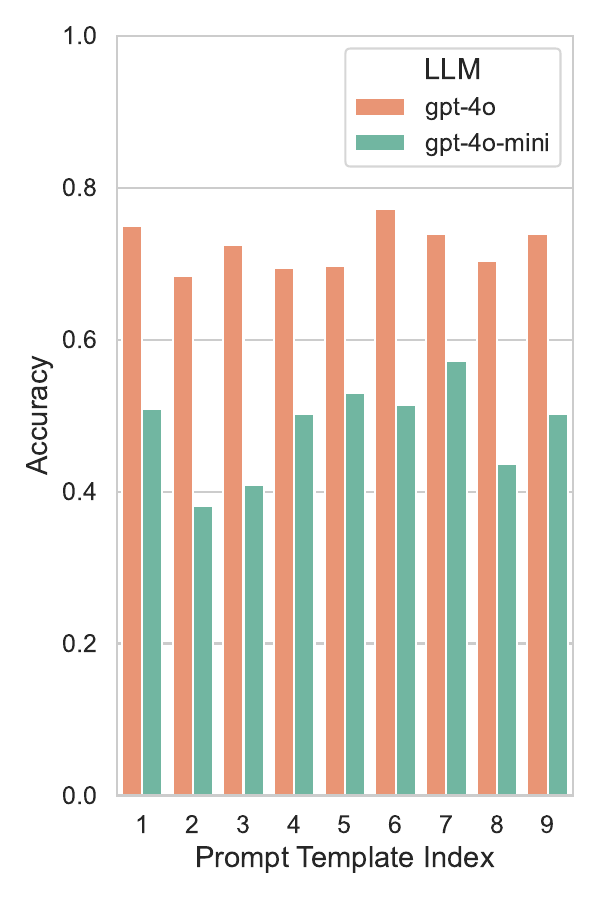}
        \caption{OCL.}
    \end{subfigure}
    \hfill    
    \begin{subfigure}[b]{0.32\linewidth}
        \centering
        \includegraphics[width=\linewidth]{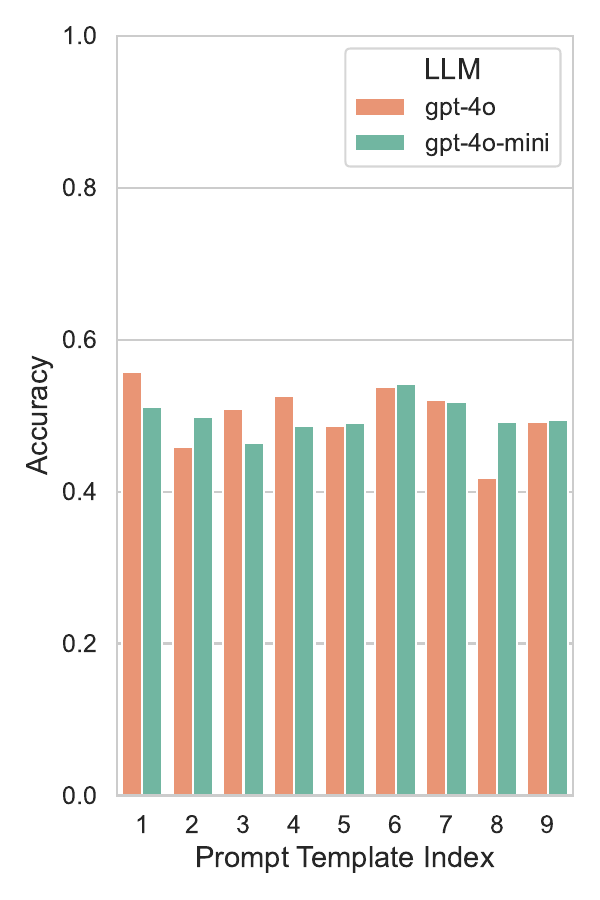}
        \caption{Alloy.}
    \end{subfigure}
    \caption{Impact of different prompting templates in the accuracy of generated code. \ins{Bars represent the performance of each LLM, with two bars shown for each prompt template.}}
    \label{fig:prompt-templates}
\end{figure}

\paragraph{Results} Figure~\ref{fig:prompt-templates} shows the relative accuracy of these prompt templates for each LLM and language.  Given that the previous Sections have discussed the effect of the target programming language and the LLM, we will focus our discussion on the impact of the prompt template in the code generation process.


\begin{table}[t]
\centering
\scriptsize
\caption{Number of tokens required by each of the prompting templates}
\begin{tabular}{@{}lccccc@{}}
\toprule
\textbf{PT} & \textbf{Domain Description} & \textbf{Domain Model} & \textbf{Explain DM} & \textbf{Code DM} & \textbf{Total} \\ 
\midrule
PT1 & 178.60 & & & & 178.60\\
PT2 & & 175.39 & & & 175.39\\
PT3 & & 293.45 & & & 293.45\\
PT4 & & 175.39 & 293.45$^*$ & & 468.84\\
PT5 & & 293.45 & 293.45$^*$ & & 586.90\\
PT6 & 178.60 & & & 293.45$^*$ & 472.05\\
PT7 & 178.60 & & & 175.39$^*$ & 353.99\\ 
PT8 & 178.60 & 175.39 & & & 353.99\\
PT9 & 178.60 & 293.45 & & & 472.05\\
\bottomrule
\end{tabular}

\smallskip
($^*$) Estimated values based on average word counts for the datasets.

\label{tab:promptingtemplates_size}
\end{table}


\paragraph{Discussion} The graphs show slight differences in correctness between different prompt templates, but there does not seem to be a clear winner or trend. For instance, in GPT-4o PT2 achieves worse results than PT1, while for GPT-4o-mini PT2 performs better than PT1 for Alloy and Python. Also, PT6 is the best for Alloy and Python, but not for OCL. Furthermore, the difference between different PTs is most noticeable for OCL, while being less relevant for Python and Alloy.

It is important to establish whether the observed differences in correctness between different prompt templates are significant or not. If not, then the choice of prompt template can be based on other factors, such as the number of tokens used by each prompt template (see Table~\ref{tab:promptingtemplates_size}). That is, if all prompt templates have comparable performance, we can use the one requiring the least amount of tokens (and thus, the cheapest). To determine whether the observed differences are significant, we will perform an \chg{statistic}{statistical} test.


Considering that accuracy is an aggregate metric that is computed from correctness, and that the correctness evaluation for each instance is represented by a binary number, a suitable statistical test is Chi-Square. It measures the discrepancy between observed and expected frequencies, as shown in Equation~\ref{eq:chi-square}, where \(\chi^2\) is the Chi-Square statistic, \(O_i\) is the observed frequency for category \(i\), \(E_i\) is the expected frequency for category \(i\), and \(n\) is the total number of categories.


\begin{equation}
\label{eq:chi-square}
\chi^2 = \sum_{i=1}^{n} \frac{(O_i - E_i)^2}{E_i}
\end{equation}

Our null hypothesis ($H_0$) considers that the differences in performance observed between prompt templates (PT) with a single LLM and a single language (as seen in Figure~\ref{fig:prompt-templates}) are not significant: there is no association between the prompt template and accuracy, therefore, the accuracy proportions are equal across all prompt templates. On the other hand, our alternative hypothesis ($H_A$) considers that there is an association between prompting templates and correctness: at least one prompting template (PT) leads to a correctness proportion (accuracy) significantly different from what we would expect under independence. Choosing a significance level of $\alpha=0.05$, when the computed P-Value is lower than the significance level, we accept our null hypothesis ($H_0$) and conclude that the differences observed between prompt templates are not significant. Otherwise, if the P-Value is \chg{equals}{equal} or greater than the significance level, we will reject our null hypothesis ($H_0$), 

\begin{table}[h!]
\centering
\scriptsize
\caption{Chi-Square Statistics for Different Models and Languages}
\begin{tabular}{@{}lccccc@{}}
\toprule
\textbf{Language} & \textbf{Model} & \textbf{Chi-Square Statistic} & \textbf{P-Value} & \textbf{Degrees of Freedom} & \textbf{Significance} \\ 
\midrule
ALLOY & GPT-4o & 21.33 & 0.0063 & 8 & \textbf{Significant} \\ 
ALLOY & GPT-4o-mini & 5.52 & 0.7004 & 8 & Not significant \\ 
OCL & GPT-4o & 12.66 & 0.1243 & 8 & Not significant \\ 
OCL & GPT-4o-mini & 43.97 & $5.76 \times 10^{-7}$ & 8 & \textbf{Highly significant} \\ 
PYTHON & GPT-4o & 4.24 & 0.8345 & 8 & Not significant \\ 
PYTHON & GPT-4o-mini & 11.36 & 0.1819 & 8 & Not significant \\ 
\bottomrule
\end{tabular}

\label{tab:chi_square_statistics}
\end{table}

Table~\ref{tab:chi_square_statistics} shows the statistical values calculated that allowed us to assess the significance of each combination of language and LLM. In total, 4 combinations do not show evidence that PT significantly affects correctness. Nevertheless, there are 2 combinations that show significant differences in performance: Alloy with GPT-4o, and OCL with GPT-4o-mini. Looking at these 2 significant combinations, it is important to identify the prompt templates that contribute to the observed differences. In particular, we observe that:
\begin{itemize}
\item For Alloy and GPT-4o, PT1, PT6 and PT7 showcase the best results, while PT3 and PT8 offer the worst.
\item For OCL and GPT-4o-mini, PT1, PT7 and PT9 offer the best results, while PT2, PT3 and PT8 offer the worst. 
\end{itemize}

In conclusion, while there is no general trend, for some combinations of languages and LLM, PT1 and PT6 achieve the best results, while PT3 and PT8 are the worst performers. There is no common distinguishable trait shared by the best PTs (PT1, PT6) and worst PTs (PT3, PT8), such as the use or formal or informal descriptions of the domain model. This implies that we cannot draw any conclusions regarding which is the best prompt template (or whether providing the context information helps or not). This \chg{proves}{disproves} our initial hypothesis\del{ wrong}.

Given that there is no clear advantage in terms of the quality of the output, PT1 can be considered the best option since it uses the least number of tokens -- which improves efficiency, sustainability, and price (if applicable).


\subsubsection{Does the way in which the coding tasks are delivered/submitted to the LLM affect the quality of the generated code?}
\label{sec:task-delivery}

\paragraph{Hypothesis} As discussed in Section \ref{subsec:prompting-and-augmentation}, when generating code for a set of integrity constraints, there are different ways to deliver these tasks. Our hypothesis is that the choice of the delivery method has an impact on the quality of the output. 

\paragraph{Experiments} In this section, we have compared \emph{batch submission}, where the code generation is required to generate all invariants in a single invocation; and \emph{isolated submission}, where a different call to the LLM is performed for each individual invariant.

\begin{figure}[ht]
    \centering
    \begin{subfigure}[b]{0.4\linewidth}
        \centering
        \includegraphics[width=\linewidth]{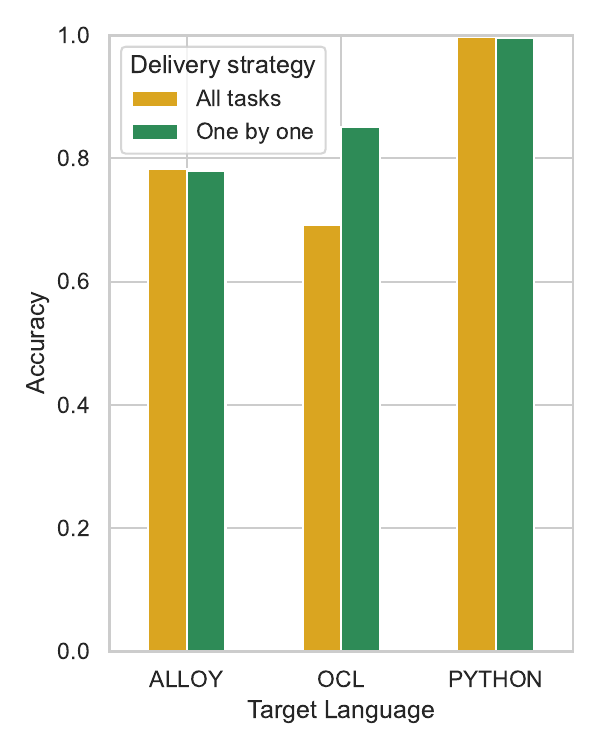}
        \caption{Well-formedness.}
    \end{subfigure}
    \hfill
    \begin{subfigure}[b]{0.4\linewidth}
        \centering
        \includegraphics[width=\linewidth]{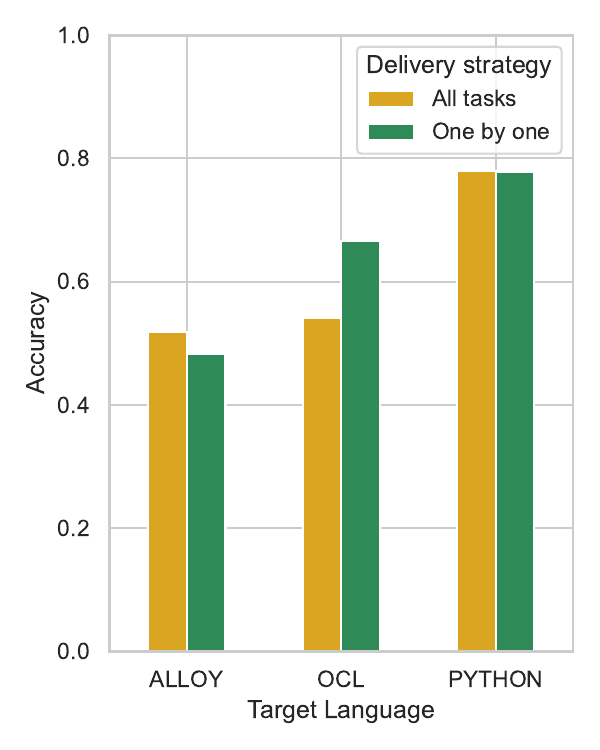}
        \caption{Correctness.}
    \end{subfigure}
    \caption{Performance of the delivery \ins{strategy} methods. \ins{Each bar represents the performance of a delivery strategy, with two bars shown for each target language. }}
    \label{fig:task-delivery-comparison}
\end{figure}

\paragraph{Results and discussion} Figure~\ref{fig:task-delivery-comparison} illustrates the accuracy of two alternative task delivery options: generating all invariants in one shot, \chg{ore}{or} generating invariants one by one. These results showcase a comparable performance of both modes, with the exception of OCL where there is a clear advantage from generating invariants individually.

However, these results only consider the correctness of the results provided in each call to the code generator and do not consider the \emph{integration} of those results for the case of isolated submission. That is, when generating invariants individually, the code generator may make assumptions about the underlying model, such as how certain information will be stored in the model. \ins{This may happen, for instance, if the prompt template being used provides a textual description of the domain model rather than a domain model, so the structure and naming of elements of the domain model needs to be assigned by the LLM.} There is no guarantee that the decisions made by each call to the code generator will be consistent among them. Thus, the code generator may assume that some information is an attribute on one call, and then that it is stored by means of an association in a subsequent call. These different assumptions may cause the resulting invariants to be incompatible, and thus impossible to integrate. As a result, considering these integration issues, the increased cost (each call to the code generator must provide the context information) and the fact the isolated submission does not yield better performance than batch submission in general, batch submission would be the preferred strategy.

Incompatibility problems can be grouped into two categories, (1) errors related to the domain model conformance, and (2) errors caused by assumptions, or inferred information, that extend or modify the original domain model. 
For example, when a LLM assumes association roles that are not defined by the domain model explicitly, these names may differ from one constraint to another, potentially causing incompatibilities between the generated invariants.
Another related example occurs when a LLM \chg{changes}{assumes the choice of a particular} an association role name \ins{when generating one constraint}: since previous and following constraints will not be aware of this \chg{change}{assumption}, navigation conflicts may appear.

\subsubsection{Do multiple attempts to generate code and/or code repair improve the quality of the generated code?}
\label{sec:code-repair}

\paragraph{Hypothesis} 
As LLMs are non-deterministic and given the same input, different queries to the LLM could lead to different outputs, running the code generation process multiple times increases the chance of producing a correct solution.
The output of the LLM can also be evaluated to detect problems in the generated code. If problems are detected, then they are reported to the LLM and a fix is requested.
Our hypothesis is that these two strategies, either isolated or combined, will help improve the quality of the generated code\footnote{Note that even if \emph{code repair} (\emph{i.e.}, requesting fixes to the LLM) is more costly because it requires invoking the LLM two more times (one for the repair itself and one to assess correctness of the repaired code), it offers an additional chance to generate correct code.}.

\paragraph{Experiments} We have applied both strategies and quantify their impact on the quality of the generated code. We have used GPT-4o and GPT-4o-mini to study the aggregated results of both LLMs.

\begin{figure}[ht]
    \centering
    \begin{subfigure}[b]{0.47\linewidth}
        \centering
        \includegraphics[width=\linewidth]{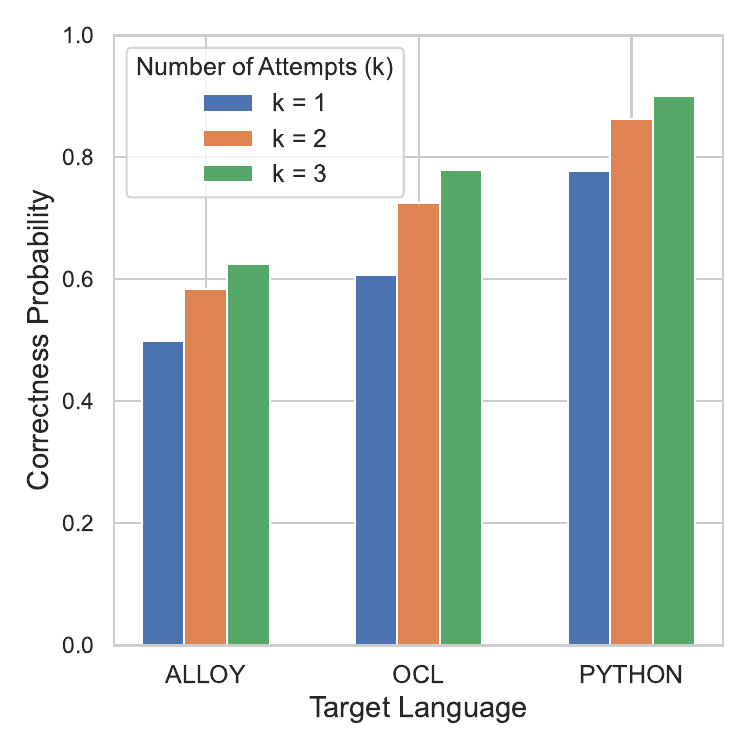}
        \caption{\chg{Pass@3.}{Correctness gain with more attempts (Pass@3).}}
    \end{subfigure}
    \hfill
    \begin{subfigure}[b]{0.47\linewidth}
        \centering
        \includegraphics[width=\linewidth]{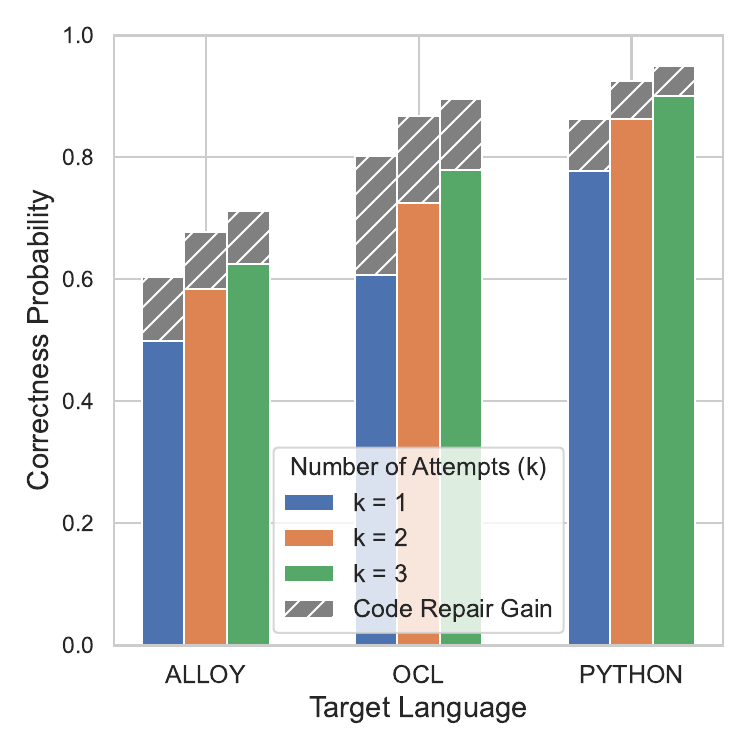}
        \caption{Pass@3 + code repair.}
    \end{subfigure}
    \caption{Correctness gain with more attempts or with code correction\ins{. Both graphs show increasing correctness probability (see Section~\ref{sec:correctness-eval-stage} and Equation~\ref{eq:pass@k}) with additional attempts. Each attempt is denoted by \emph{k} and represented with a bar, where the first attempt correspond to \emph{k = 1}. The second graph additionally illustrates the effect of code repair. The gray region on the top of each attempt bar represents the additional gain achieved through repair.}}
    \label{fig:code-repair}
\end{figure}

\paragraph{Results and discussion} Figure~\ref{fig:code-repair} illustrates the effects of allowing \emph{k} code generation attempts (where 1$\leq$\emph{k}$\leq$3) for the same task as well as the inclusion of a round of code repair for incorrect code. In Fig.~\ref{fig:code-repair}(a) we see the correctness gains achieved by increasing the number of attempts (k). Increasing the number of attempts has a linear increase in terms of costs, and it increases the probability of obtaining correct generated code. However, as expected, there are diminishing returns: as we increase the number of attempts, the percentage of correct tasks also increases but each attempt provides smaller improvements.

Fig.~\ref{fig:code-repair}(b) shows the effects of including code repair in the code generation pipeline. Code repair always increases the likelihood of producing correct code, as it is only invoked for those cases where the initially generated code was incorrect. There is an increase of 10-20\% in the number of correctly generated tasks when applying a single round of code repair.

Moreover, code repair can be combined with multiple (k) attempts, which implies applying code repair to each incorrect generated piece of code. This combination achieves the best results. The only disadvantage is that it requires more  calls to the LLM, thus increasing the cost of the code generation process.

\subsection{Discussion and takeaways}
\label{sec:discussionAndTakeaways}

\ins{Our findings highlight several important aspects of LLM performance in code generation across different programming languages. First, the target language plays a crucial role in determining the appropriate model choice. Performance can vary significantly, especially for low-resource languages, where selecting a model with prior exposure to the language becomes critical.}

\ins{When an LLM is already knowledgeable of the target language, the influence of prompt engineering appears to diminish. In such cases, most prompts lead to similar results, suggesting that linguistic familiarity can compensate for less optimized prompt design.}

\ins{Moreover, we observe that including multiple constraints within a single prompt, provided they belong to the same domain, tends to produce better and more consistent outputs. Providing constraints to the LLM separately may lead to conflicts or incompatible interpretations by the model.}

\ins{Finally, re-prompting a task multiple times (\emph{i.e.}, multiple attempts) generally leads to higher success rates compared to asking the model to repair an incorrect output after a failed attempt. This indicates that regenerating solutions from scratch may be effective. Given that code repair is sometimes successful in fixing problems in the generated code, a combination of multiple attempts with code repair is the best performing strategy, although it is also the one that requires the most resources.}

\ins{Table~\ref{tab:practitioner_guidelines} collects a series of concrete guidelines for practitioners, summarizing the key takeaways gathered from our findings.}

\begin{table}[t]
\centering
\caption{\ins{Practical guidelines for practitioners}}
\label{tab:practitioner_guidelines}
\begin{tabularx}{\linewidth}{@{}c X X@{}}
\toprule
\textbf{ID} & \textbf{Guideline} & \textbf{Rationale / Practical Insight} \\
\midrule
G1 &
\ins{Prioritize selecting a model that is well aligned with the target language, particularly for low-resource or less common languages.} &
\ins{Model–language alignment has a larger impact on output quality than prompt tuning when language support is limited.} \\

\addlinespace
G2 &
\ins{When the model already supports the target language well, avoid extensive prompt engineering; instead, rely on basic validation and multiple attempts.} &
\ins{Over-engineering prompts yields diminishing returns when baseline language competence is strong.} \\

\addlinespace
G3 &
\ins{Specify multiple constraints from the same domain within a single prompt rather than across multiple prompts.} &
\ins{Consolidating related constraints reduces the risk of incompatible or contradictory generations.} \\

\addlinespace
G4 &
\ins{If resources make it viable, use multiple code generation attempts and code repair.}&
\ins{Multiple attempts and code repair can both improve code correctness and are complementary to each other.} \\
\bottomrule
\end{tabularx}
\end{table}

\subsection{Threats to validity}
\label{sec:threats}

In this subsection, we discuss the potential threats to validity that may affect our experimental design and the interpretation of the results.

Our experiments have considered code generation tasks for three target languages: OCL, Alloy and Python. OCL and Alloy are two usual formal notations for describing integrity constraints in systems. The results we have obtained may differ for other general-purpose languages and DSLs, as the performance of LLMs depends on whether the target language (or similar ones) is present in the training dataset. To mitigate this, future work could include a broader range of target languages, particularly underrepresented or low-resource DSLs, and explore model fine-tuning or prompting strategies tailored to such languages.

Also, our experiments have used a fixed set of LLMs for code generation: GPT-4o, GPT-4o-mini, DeepSeek Coder 6.7b and Llama 3.1. Other LLMs may achieve different performance in the task under study and be affected differently by factors such as prompt templates, batch delivery, code repair, and multiple attempts. Precisely, the proposed framework has been built to support the evaluation of different LLMs and different code generation strategies. It is worth noting that, in our preliminary experiments, our initial idea was to study a larger set of LLMs in Ollama (including Mistral 7b, Code Llama 7b, Qwen 2.5 Coder 7b, and Granite Code 8b). Unfortunately, we had to discard them for several reasons, including incompatibility with our hardware (NVIDIA RTX A5000 with 24GB of VRAM), frequent freezing, or excessively long response times ranging from 5 to 10 minutes.

Given the large number of tasks and alternative code generation strategies, we have relied on  LLM-as-a-judge to provide an automatic assessment of the correctness of generated code. In Section~\ref{sec:manual-automatic}, we have measured its accuracy compared to manual code evaluation in 1,512 tasks. While being less precise than manual evaluation, it is becoming a well-accepted alternative as it offers a reasonable compromise that avoids subjectivity, variability and enables the large-scale evaluation and comparison of multiple code generation strategies~\cite{gu_survey_2025}.

Finally, our experiments use two datasets (Section~\ref{sec:datasets}) obtained and adapted from the literature, which ensures that our evaluation is based on publicly available and widely accepted sources. While this adds transparency and reproducibility, it also introduces external biases. Our results may be related to the domain information and the type of constraints used in these datasets. Some constraints appearing in those datasets may be easier to encode in some target constraint language, leading to unfair comparison. For example, constraints referring to dates and times may be easier to express in a general-purpose language like Python (which has native datatypes to handle these concepts) than in DSLs like Alloy or OCL (which do not include a native Date datatype). \ins{Moreover, our experiments have used textual descriptions of the domain models in our datasets that are generated by an LLM. As a result, these synthetic descriptions may differ from human-created domain descriptions, for instance, being more concise, less ambiguous or more predictable. This may affect external validity, limiting our ability to generalize these results to other domains with human-created textual specifications.}

\section{Related work}
\label{sec:related-work}


Abukhalaf et al. \cite{abukhalaf_codex_2023} investigated the generation of Object Constraint Language (OCL) constraints using Codex, emphasizing the importance of combining UML models with natural language specifications in prompts to enhance correctness. Their work involved the automatic evaluation of well-formedness using USE-OCL \cite{gogolla_use_2007} (a tool for modeling, validating, and testing OCL constraints within UML models) and the manual assessment of correctness. Additionally, they developed a dataset comprising 15 UML models and 114 OCL constraints, each paired with its respective natural language specification, to support and validate their findings.

Another prompt augmentation approach reduces the UML model used as context and instead includes only the relevant subset of classes in the prompt with the UML model. OCL constraints were generated using GPT-4, with automatic evaluation of well-formedness conducted via USE-OCL execution and manual evaluation of correctness \cite{abukhalaf_pathocl_2024}. However, there is no mention of any specific technique for correctness evaluation or functional correctness assertion through unit test cases, such as a test case scenario involving model instantiation in USE-OCL.

A locally fine-tuned language model demonstrates similar syntactic accuracy to GPT-4 Turbo while surpassing it in semantic similarity for OCL generation \cite{pan_generative_2024}. In order to be able to fine-tune the model, a dataset with 52 models and 369 OCL constraints was created with synthetic natural language specifications. They used executions of Eclipse EMF and OCL plugins to assess well-formedness, and an exact match to assess for correctness approach based on cosine similarity.

Hong et al. \cite{hong_effectiveness_2025} conducted a qualitative study evaluating the effectiveness of using LLMs for generating Alloy formulas. They focused on the analysis of the Alloy code generation of graph and binary relations basic properties. Specifically, they used an Alloy specification, and prompt the LLM to complete the specification according to the coding task. As a result, the authors did not aim to generate a complete Alloy specification, including the modeling aspect, nor did they explore how to effectively address such scenarios. In contrast, our work focuses specifically on constraint generation within this context. Their research complements our efforts by examining other scenarios for Alloy code generation. We notice that they relied on manual executions of the Alloy Analyzer to evaluate well-formedness and conducted manual checks to ensure functional correctness. They assert that their work represents the first study leveraging LLMs to address traditional synthesis problems using Alloy.

Testing plays a crucial role in assessing the correctness of generated Alloy code. The AUnit testing framework \cite{sullivan_towards_2014} provides foundational concepts for unit testing declarative models written in Alloy, including definitions for test cases, coverage, and coverage criteria. While the initial AUnit framework laid the groundwork, the tool proposed \cite{sullivan_aunit_2018} primarily focused on integrating AUnit as an extension within the Alloy Analyzer, which does not fully automate the testing process in a way that directly supports the easy and automated validation of generated Alloy code. Nevertheless, the core concepts of the AUnit framework, such as its definition of declarative test cases (a valuation and an Alloy command) and model coverage criteria, remain highly relevant and could be instrumental in automating correctness validation for generated code, even if further development is required to achieve full automation.

Finally, when the program fails to compile or it does not pass any of the tests, the feedback and the faulty code can be used as input to prompt the LLM to explain the problem and to fix the code \cite{olausson_is_2024}.

Fine-tuning re-trains an LLM for a new task by adjusting its AI model parameters. This process requires an extensive dataset for this retraining process, which in the case of code generation would be a large collection of code samples. However, in some cases this is considered impractical, e.g., due to the data scarcity of DSLs as well as the high computational cost of the re-training. In contrast, parameter-efficient fine-tuning achieves similar results to fine-tuning while keeping most of the language model unchanged \cite{hou_large_2024}: it updates only a small set of parameters, making the process more efficient and requiring fewer computational resources. This technique has been proven more effective than full fine-tuning, in-context learning, and retrieval-augmented generation (RAG) with LLMs targeting code generation \cite{weyssow_exploring_2025, hou_large_2024}.
\section{Conclusions}
\label{sec:conclusions}

In the context of low-resource programming languages, code generation using LLMs faces significant challenges. The lack of examples in the LLM's training datasets causes problems in terms of the well-formedness and correctness of the generated code. In this paper, we have proposed a framework to evaluate the capabilities of LLMs generating code from textual requirements in terms of both well-formedness and correctness. The proposed framework is flexible, allowing the comparison of different LLMs, code generation prompts, and the use of different quality evaluation tools for code. 

We have decided to study code generation for a family of DSLs: constraint and query languages. Using our framework, we have evaluated $\approx$100,000 code generation tasks for Alloy, OCL and Python using different code generation strategies. 
Given the number of experiments, the use of automated evaluation (rather than manual evaluation) has been a necessity. Experiments show that the target language and LLM of choice are the most relevant factors affecting the quality of generated code. Enabling code repair and multiple code generation attempts improve the quality of generated code, while the prompt template and task delivery mode have a lesser impact. While some of these results are particular to constraint and query languages, some lessons learned can also be relevant to other DSLs. For reproducibility purposes, we provide all the software artifacts used in an open source repository~\cite{repo}.

Regarding future work, we would like to integrate in our framework and test the impacts of retrieval-augmented generation (RAG) techniques to retrieve the relevant pieces of the domain model that are relevant for a particular piece of code. 
Moreover, we would like to study how to make the well-formedness and correctness checkers available as tools to the LLMs, \emph{e.g.}, using approaches like Model Context Protocol (MCP) \cite{anthropic_introducing_2024}, and to assess whether code generation can improve with the availability of such tools to the LLM. We will look into how to integrate fine-tuning into our assessment framework, \emph{e.g.}, how to use the examples of generated code during the evaluation as data for the fine-tuning process. Finally, we would like to evaluate our framework with different datasets and languages.

\section*{Acknowledgement}
This project has been funded by the Spanish Ministry of Science, Innovation, and Universities under contract PID2021-125527NB-I00 and PID2023-147592OB-I00 (project SE4GenAI); the research network RED2022-134647-T (MCIN/AEI/10.13039/501100011033, AI4SE network); University of Málaga under project JA.B1-17 PPRO-B1-2023-037 and a doctoral grant from Universitat Oberta de Catalunya.


\appendix

\section{Example of a complete specification}

\ins{In this appendix, we provide a complete example of a coding task. We describe three elements: the domain model, the target constraint (which is defined over the previous domain) and the generated code in the three target languages (OCL, Alloy and Python).} 

\ins{We show two alternative characterizations of the domain model: as a textual description in natural language or as a class diagram in PlantUML. Our framework will use one description or the other depending on the prompt template selected by the user. The sample generated code shown in this Appendix has been generated using the prompting template \emph{1} or \emph{PT1} (generate code from a textual domain description).}

\paragraph{\ins{Domain description}} \ins{The textual domain description is the following:}

\begin{quote}
{\ins{We have accidents, which may be of a specific type for example, a rear end collision. Each accident occurs on a roadway and may involve multiple victims, with a recorded number of fatal victims.} 

\ins{Vehicles are captured. Some particular kind of vehicles are those that are traveling, and more specifically, we also distinguish those involved in accidents, are classified as crashed vehicles. A crashed vehicle is associated with a single accident and the roadway where the accident occurred. Each accident involves one or more such vehicles.} 

\ins{The system needs to keep track of people. There are three primary subtypes according to the system: living persons, deceased persons and travelers. Travelers can be further categorized into drivers, passengers and victims.}

\ins{Each Travel instance represents a trip, consisting of one vehicle and one or more travelers. This allows us to trace who was traveling in which vehicle at the time of the incident.}}
\end{quote}

\begin{figure}[h]
\centering
\includegraphics[width=0.9\columnwidth]{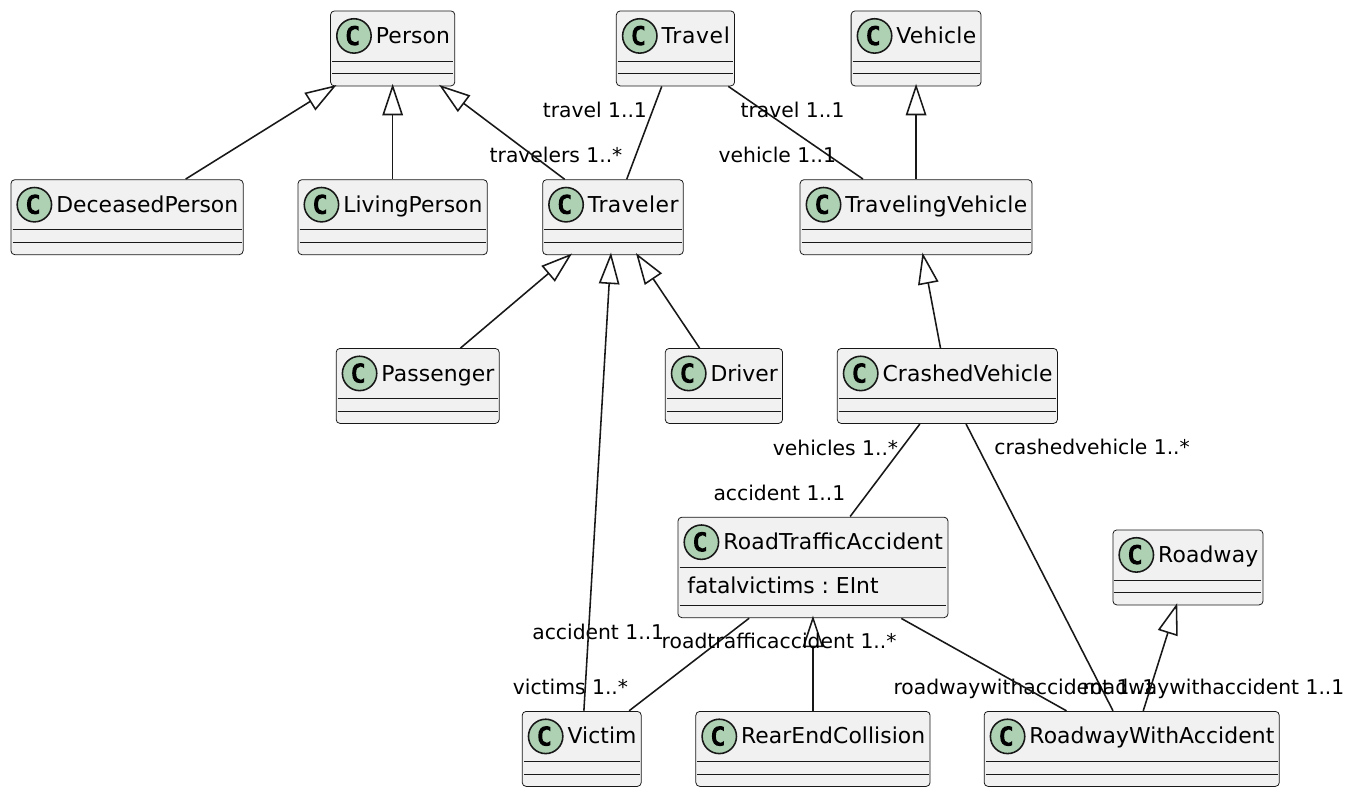}
\caption{\ins{Domain model in PlantUML.}}
\label{fig:domain-model}
\end{figure}

\paragraph{\ins{Domain model in PlantUML}} \ins{Figure~\ref{fig:domain-model} shows the PlantUML class diagram that captures the previously described domain.}

\paragraph{\ins{Coding task}} \ins{The following is the description of the target constraint in natural language:} 

\begin{quote}
\ins{A rear-end collision must involve exactly two vehicles.}
\end{quote}

\paragraph{\ins{Generated code}} \ins{The following listings show the code generated in Python, OCL, and Alloy for the previous coding task considering the textual domain description. Notice that the Python code is a boolean function that checks whether the property holds, the OCL code formalizes the constraint as an invariant and the Alloy code expresses it as a fact.}

\begin{lstlisting}[style=python]
def check_rear_end_collision_constraint(accidents):
    for accident in accidents:
        if accident['type'] == 'rear end collision' and len(accident['crashed_vehicles']) != 2:
            return False

    return True
\end{lstlisting}

\begin{lstlisting}[style=ocl]
context Accident inv: 
    self.type = 'rear-end collision' implies self.crashedVehicles->size() = 2
\end{lstlisting}

\begin{lstlisting}[style=alloy]
fact RearEndCollisionInvolvesTwoVehicles { 
    all a: Accident | a.type = RearEndCollision implies #a.crashedVehicles = 2 
}
\end{lstlisting}








\bibliographystyle{elsarticle-num} 
\bibliography{references}

\end{document}